\documentclass[amsmath,amssymb,nofootinbib,notitlepage,superscriptaddress,twocolumn]{revtex4-1}
\pdfoutput=1

\usepackage[bottom]{footmisc}
\usepackage[colorlinks=true]{hyperref}
\usepackage[raggedright]{subfigure}
\usepackage{lipsum, babel}
\usepackage{comment}
\usepackage{slashbox}
\usepackage{amssymb}
\usepackage{cancel}
\usepackage{pifont}
\usepackage{soul}
\newcommand{\checks}{\ding{51}}%
\newcommand{\x}{\ding{55}}%
\usepackage{mathtools,amsmath,amssymb,mathrsfs,color,esint}
\usepackage{array}
\newcolumntype{C}[1]{>{\centering\arraybackslash}p{#1}}

\newcommand{\sign}{\text{sign}}
\newcommand{\EVPA}{\text{EVPA}}

\begin{document}

\title{Polarized Image of Equatorial Emission in the Kerr Geometry}

\author{Zachary Gelles}
\email{zgelles@college.harvard.edu}
\affiliation{Center for Astrophysics $\vert$ Harvard \& Smithsonian, 60 Garden Street, Cambridge, MA 02138, USA}
\affiliation{Black Hole Initiative at Harvard University, 20 Garden Street, Cambridge, MA 02138, USA}
\author{Elizabeth Himwich}
\email{himwich@g.harvard.edu}
\affiliation{Center for the Fundamental Laws of Nature, Harvard University, Cambridge, MA 02138, USA}
\affiliation{Black Hole Initiative at Harvard University, 20 Garden Street, Cambridge, MA 02138, USA}
\author{Michael~D.~Johnson}
\email{mjohnson@cfa.harvard.edu}
\affiliation{Center for Astrophysics $\vert$ Harvard \& Smithsonian, 60 Garden Street, Cambridge, MA 02138, USA}
\affiliation{Black Hole Initiative at Harvard University, 20 Garden Street, Cambridge, MA 02138, USA}
\author{Daniel~C.~M.~Palumbo}
\email{daniel.palumbo@cfa.harvard.edu}
\affiliation{Center for Astrophysics $\vert$ Harvard \& Smithsonian, 60 Garden Street, Cambridge, MA 02138, USA}
\affiliation{Black Hole Initiative at Harvard University, 20 Garden Street, Cambridge, MA 02138, USA}

\begin{abstract}

We develop a simple toy model for polarized images of synchrotron emission from an equatorial source around a Kerr black hole by using a semi-analytic solution of the null geodesic equation and conservation of the Penrose-Walker constant.  
Our model is an extension of \cite{Narayan_2020}, which presented results for a Schwarzschild black hole, including a fully analytic approximation. Our model includes an arbitrary observer inclination, black hole spin, local boost, and local magnetic field configuration. We study the geometric effects of black hole spin on photon parallel transport and isolate these effects from the complicated combination of relativistic, gravitational, and electromagnetic processes in the emission region. 
Expanding in $1/r_{\rm s}$, we find an analytic approximation, consistent with previous work, for the geometric effect of spin on observed face-on polarization rotation in the direct image: $\Delta {\rm EVPA} \sim -2a/r_{\rm s}^2$, where $a$ is the black hole spin and $r_{\rm s}$ is the emission radius. We further show that spin introduces an order unity effect on face-on subimages: $\Delta {\rm EVPA} \sim  \pm a/\sqrt{27}$. 
We also use our toy model to analyze polarization ``loops" observed during flares of orbiting hotspots. Our model provides insight into polarimetric simulations and observations of black holes such as those made by the EHT and GRAVITY.

\end{abstract}

\maketitle

\section{Introduction}

The first polarized images of the black hole M87*, which reveal a bright ring of emission with twisting polarization pattern, have recently been released by the Event Horizon Telescope (EHT) collaboration~\cite{PaperI,PaperII,PaperIII,PaperIV,PaperV,PaperVI,PaperVII,PaperVIII}. The polarization structure in black hole images depends on propagation effects, plasma emission, magnetic field geometry, and spacetime curvature. Simulations of polarized emission are an important tool that have been used to study astrophysical and geometric properties of black hole accretion flows, and as an aid for interpreting observations \cite{Connors:1980,Bromley:2001,Broderick:2003bg,Broderick:2004,Broderick:2005,Broderick:2006,Fish:2009,Zamaninasab:2010,Penna_2010,Li:2009,Shcherbakov:2011,Shcherbakov:2012,Dexter_2016,Moscibrodzka:2017,Gold:2017,Marin:2018,Jimenez:2018,Moscibrodzka:2018,Palumbo_2020,Moscibrodzka:2020,Moscibrodzka:2021}. Detailed simulations that simultaneously incorporate astrophysical and relativistic effects are physically realistic but are generally computationally expensive. Furthermore, disentangling astrophysical and relativistic effects in these models can be challenging.

Toy models offer an efficient pathway to decouple and characterize different effects on black hole images over a broad range of simplified emission configurations. An exact description of polarized images from emission around black holes requires a numerical solution of the geodesic equation, even in the spherically symmetric Schwarzschild geometry (see, e.g., \cite{Chen_1991}). The description of polarized images in both Schwarzschild and Kerr is greatly simplified by conservation of the Penrose-Walker constant \cite{Walker_Penrose_1970}. Recently, a simplified toy model of synchrotron emission around a Schwarzschild black hole was presented in \cite{Narayan_2020}, which took advantage of an additional simplification: an approximation developed by Beloborodov \cite{Beloborodov_2002} with which the polarized image can be computed analytically. Here, we extend the toy model of \cite{Narayan_2020} using tools developed in \cite{Carter_1968,Gralla2020a,Gralla2020b,Himwich2020} to include the effects of spin by generalizing to the Kerr geometry, for which a Beloborodov-like approximation is not available. 

Our model consists of an equatorial ring of magnetized fluid orbiting a Kerr black hole. The images of axisymmetric rings of radiating fluid are described by analyzing the local frame of an emitting point source in the Kerr geometry. The semi-analytic description of an unpolarized image from such an emitter has a long history going back to the 1970s with Cunningham and Bardeen, \cite{Bardeen1972,Cunningham1973} and has recently been discussed in specific contexts such as the near-horizon-extreme-Kerr emission \cite{Gralla2018,Gates2020,Gates:2020els}. The polarized image has also been treated analytically for the high-spin case in \cite{Gates:2018hub}. 

A semi-analytic treatment of the geometric effects of spin on the polarized image of orbiting geodesic rings at arbitrary emission radius and inclination was performed in the seminal work of Connors, Piran, \& Stark \cite{Connors:1980}, as well as the PhD thesis of Eric Agol \cite{Agol:1997}. 
Our model generalizes this work to semi-analytically compute the polarized image given an arbitrary spin, emission radius, magnetic field geometry, equatorial fluid velocity, and observer inclination.

Using conservation of the Penrose-Walker constant \cite{Walker_Penrose_1970}, our model generates predictions for linear polarization angle and relative polarized intensity, providing insight into how the accretion flow and spacetime geometry affect the polarized image. In this paper, we provide illustrative examples of polarized images for a variety of physical configurations. Our model includes the image of direct emission as well the corresponding lensed indirect emission, or ``subimages." Though we consider only emission from a single radius, the image of a disk with finite radial extent can be modelled by simply summing contributions from individual radii; this was performed by \cite{Agol:1997}, which noted that the sum gave rise to a net depolarization effect, since adding polarization vectors at different angles reduces the polarization. An analogous depolarization can arise from summing contributions of direct and indirect emission, as seen in simulations \cite{Jimenez-Rosales:2021ytz}.

For face-on black holes, we find an analytic expression for the effect of spin on polarization rotation, which is subleading in $1/r_{\rm s}$ in the direct image ($\Delta {\rm EVPA} \sim -2a/r_{\rm s}^2$) but order unity in subimages ($\Delta {\rm EVPA} \sim  a/\sqrt{27}$),  where $a$ is the black hole spin and $r_{\rm s}$ is the emission radius.
The specific effects of frame-dragging on the rotation of the polarization plane have been studied extensively in the past and have been interpreted as a gravitational analogue of Faraday rotation \cite{Balazs_1958,Pineault_1977,Fayos_1982,Brodutch_2011,Ishihara_1988,Nouri_1999}. In this work, we use our model to re-derive earlier results using simple techniques that avoid ambiguities about local reference frames. We further build on prior work by examining the effects of parallel transport on photon ring subimages.

An important EHT observing target is the black hole Sgr~A* at the center of our own galaxy. The polarization of Sgr~A* has significant time variability in both submillimeter \cite{Marrone_2006,Johnson:2015,Bower:2018} and near-infrared \cite{Eckart:2006,Trippe:2007,Gravity_2018,Gravity_2020} observations, with particularly rapid variability in near-infrared flares. The flares are likely born out of various plasma and MHD effects such as magnetic reconnection \cite{Dodds_2010,Ripperda_2020}. Simulations are a powerful tool used to understand these polarized flares and have a long history also going back to the 1970s, including the work of \cite{Connors:1980}, which presents time-dependent EVPA in the direct emission from an orbiting hotspot. More advanced recent simulation studies include \cite{Broderick:2005,Broderick:2006,Fish:2009,Zamaninasab:2010}, which attribute certain time-varying features to emission from a localized orbiting hotspot. These features may be manifested as loops in the linear Stokes $Q,U$ polarization. Our toy semi-analytic model, which can isolate geometric effects from the simulated astrophysical processes, complements these efforts. Our model also isolates the effects of individual subimages and the ways in which they affect observed polarization patterns.

Using analytic results in the Kerr geometry, we can model the effects of spin on the image of hotspot emission and generate images of $Q,U$ time variation in direct and indirect images that reproduce observed polarization loops. Our model can be directly compared to simulated and observed values for the case of near-infrared flares, in which the influence of Faraday effects, absorption, and background emission are insignificant \cite{Narayan_2020}. 

In particular, using infrared interferometry, the GRAVITY collaboration has recently reported the first resolved centroid motion and polarization during flares of Sgr~A* \cite{Gravity_2018}. Both the centroid and polarization traced loops over time, which were interpreted using a model of an orbiting equatorial hotspot \cite{Gravity_2018,Gravity_2020}. In simulations of motion of a small Gaussian hotspot, \cite{Gravity_2018} (see Appendix D therein) found that the presence of a single polarization loop in $Q,U$, in which the orbital period in polarization is the same as the orbital period of the hotspot, is a signature of magnetic fields perpendicular to the orbital plane (i.e. vertical). In contrast, for a toroidal field (i.e. equatorial) they found that the orbital period of polarization is half that of the hotspot, corresponding to two loops in $Q,U$. The $Q,U$ data for the observed July 28, 2018 flare were consistent with a single polarization loop observed at low inclination, as in a poloidal field configuration \cite{Gravity_2018}. Follow-up work in \cite{Gravity_2020} considered the July 28 flare in further detail, and compared observations to simulations of a Gaussian hotspot as well as a simplified non-relativistic analytic model of a point emitter, finding that a single loop in $Q,U$ arises only from fields with a nonzero vertical component, with the best fit to the July 28 flare having a vertical plus azimuthal field. We substantiate the same claims in \cite{Gravity_2018,Gravity_2020} about single (double) loops arising from vertical (equatorial) magnetic fields using analytic results in the Kerr geometry, providing additional physical insight into GRAVITY observations and the general observational distinction between vertical and equatorial fields.

Our work can also be further developed to future extensions that include non-equatorial emission, and to studies of circular polarization (Stokes $V$) \cite{Bower:2002,Huang:2008,Shcherbakov:2011,Munoz:2012,Gold:2017,Ricarte:2021frd,Moscibrodzka:2021}. These extensions will be useful in comparisons to general-relativistic magnetohydrodynamic (GRMHD) simulations and will further our understanding of the extent to which fluid configuration and magnetic field geometries can be inferred using polarized images. Comparing our model to ray-traced GRMHD simulations will also provide insight into whether Faraday effects prevent such inferences.

This paper is organized as follows. Section \ref{sec:FluidModel} describes our simple model for fluid orbiting a Kerr black hole.
Section \ref{sec:Image} reviews the calculation of the observed appearance of polarized emission around a Kerr black hole.
Section \ref{sec:Visualizations} presents and discusses sample polarized images produced by our model for a variety of magnetic fields, observer inclinations, and fluid configurations. Section \ref{sec:EVPA} considers the observed EVPA for black holes viewed on-axis, quantifying the effects of spin on EVPA. Section \ref{sec:Hotspots} considers the application of our model to the polarized images of orbiting hotspots and provides additional details of  comparisons with \cite{Gravity_2018,Gravity_2020}. Section \ref{sec:Summary} gives a brief summary of our conclusions and directions for future investigation. We present explicit details of the orbiting fluid model in Appendix \ref{app:OrbiterDetails}, details of the semi-analytic solution of the geodesic equation in Appendix \ref{app:AnalyticDetails}, a general discussion of image symmetries in Appendix \ref{app:symmetries}, and details of cusp formation in direct emission $Q,U$ loops in Appendix \ref{app:cuspangle}. 

\section{Orbiting Fluid Model} \label{sec:FluidModel}

This section introduces our model of accreting fluid around a Kerr black hole using an orbiting emitter. 
\subsection{Circular Orbiting Emitter in Kerr} \label{subsec:CircularEmitter}

The Kerr line element for a black hole of mass $M$ and angular momentum $J=aM$ in Boyer-Lindquist coordinates $(t,r,\theta,\phi)$ is \cite{Kerr1963,Chandrasekhar:1985}
\begin{equation}
ds^2=-\frac{\Delta\Sigma}{\Xi}dt^2 +\frac{\Sigma}{\Delta} d r^2+\Sigma d\theta^2+\frac{\Xi\sin^2{\theta}}{\Sigma}\left[d \phi- \omega d t\right]^2, \\
\end{equation}
where
\begin{gather}
	\Delta(r)=r^2-2Mr+a^2, \quad \Sigma(r,\theta)=r^2+a^2\cos^2{\theta}, \nonumber \\
        \omega = \frac{2 a M r}{\Xi}, \quad \Xi = (r^2 + a^2)^2 - \Delta a^2 \sin^2\theta.
\end{gather}
Note that $\omega$ is the angular momentum of the zero-angular-momentum-observer (ZAMO), which vanishes in the Schwarzschild limit $a \to 0$. 

Consider a point source at radius $r_{\rm s}$ on an equatorial $(\theta_{\rm s} = \frac{\pi}{2})$ circular orbit of zero angular momentum with angular velocity $\omega_{\rm s} = \omega(r_{\rm s}, \theta_{\rm s} = \frac{\pi}{2})$.\footnote{Angular velocity is defined relative to the asymptotic rest frame by $\frac{d\phi}{dt} = \frac{u^{\phi}}{u^t}$, where $u^{\mu}$ is the four-velocity the emitting source.} The tetrad that describes the ZAMO local rest frame consists of its four-velocity $u^{\mu} = e^{\mu}_{(t)}$ ($u^{\mu}u_{\mu} = -1$) and three orthogonal unit spacelike vectors:
\begin{equation}\label{eq:zamotetrad}
  \begin{aligned}
    e_{(t)} &= \frac{1}{r_{\rm s}}\sqrt{\frac{\Xi_{\rm s}}{\Delta_{\rm s}}}(\partial_t + \omega_{\rm s} \partial_{\phi}), \\
    e_{(r)} &= \frac{1}{r_{\rm s}}\sqrt{\Delta_{\rm s}} \partial_r, \\
    e_{(\phi)} &= \frac{r_{\rm s}}{\sqrt{\Xi_{\rm s}}}\partial_{\phi}, \\
    e_{(\theta)} &= - \frac{1}{r_{\rm s}} \partial_{\theta}. \\
  \end{aligned}
\end{equation}
Note that the minus sign in the last line implies that the local $(\hat{\theta})$ and Boyer-Lindquist $\hat{\theta}$ are anti-aligned. Here and below, the subscript $s$ denotes a quantity evaluated at the source $r_{\rm s}, \theta_{\rm s} = \frac{\pi}{2}$. The local orthonormal frame has the flat Minkowski metric $\eta^{(a)(b)}$, and the frame components of four-vectors are given by
\begin{equation} \label{eq:tetradvector}
  V^{(a)} = \eta^{(a)(b)}e_{(b)}^{\mu}V_{\mu}.
\end{equation}
The orientation is such that $(\hat{x}, \hat{y}, \hat{z}) \leftrightarrow \left((\hat{r}), (\hat{\phi}), (\hat{\theta})\right)$. The tetrad is given explicitly as a matrix in Appendix \ref{app:OrbiterDetails}. 

\subsection{Boosted Emitter} \label{subsec:Boost}

From the local orthonormal ZAMO frame, consider boosting the emitter in the $(r),(\phi)$ plane with velocity
\begin{equation} \label{eq:betaboost}
\vec{\beta} = \beta_v\left(\cos\chi\,(\hat{r})+\sin\chi\,(\hat{\phi})\right).
\end{equation}
Vectors are boosted via a Lorentz transformation $\Lambda^{(a)}_{\ \ (b)}$. To transform from the ZAMO frame to the boosted orthonormal frame, denoted by primed quantities, take
\begin{equation} \label{eq:boostedframe}
\begin{aligned}
V'^{(a)} &= \Lambda^{(a)}_{\ \ (b)} V^{(b)} \\
&= \Lambda^{(a)}_{\ \ (b)} \eta^{(b)(c)} e_{(c)}^{\mu} V_{\mu},
\end{aligned}
\end{equation}
using equation \eqref{eq:tetradvector}. This defines a new tetrad
\begin{equation} \label{eq:boostedtetrad}
\begin{aligned}
e'^{\mu}_{(d)} = \eta_{(d)(a)}\Lambda^{(a)}_{\ \ (b)} \eta^{(b)(c)} e_{(c)}^{\mu} = \Lambda_{(d)}^{\ \ (c)} e_{(c)}^{\mu},
\end{aligned}
\end{equation} 
with components given by matrix multiplication (recall that for $\Lambda = \Lambda^{(a)}_{\ \ (b)}$, the inverse matrix $ \Lambda^{-1} = \Lambda_{(a)}^{\ \ (b)}$). For concreteness, the explicit tetrad components are given in Appendix \ref{app:OrbiterDetails}. An inverse transformation from the local frame to the vector in Kerr is given by 
\begin{equation} \label{eq:invtetrad}
V^{\mu} = e'^{\mu}_{(a)}V'^{(a)}.
\end{equation}
The next section uses the local emitter motion to compute its polarized image seen by a distant observer.

\section{Image Intensity and Polarization} \label{sec:Image}

This section describes how to compute the observed location and polarization of the direct image of an axisymmetric equatorial disk of emitting matter at radius $r_{\rm s}$ around a Kerr black hole with arbitrary spin $a$, local magnetic field $\vec{B}$, and observer inclination $\theta_o$. We briefly summarize the important steps in the calculation, details of which are included in the following  subsections: 

\begin{enumerate}
\item Subsections \ref{subsec:LightProp} and \ref{subsec:BardeenCoords}: For photons emitted by a source at position $r_{\rm s}, \theta_{\rm s} = \frac{\pi}{2}$, we find the arrival position on the screen (and corresponding conserved quantities) by solving the geodesic connecting the source $r_{\rm s}, \theta_{\rm s} = \frac{\pi}{2}$ and observer. 
\item Subsection~\ref{subsec:PolLocalFrame}: The photon conserved quantities give its initial momentum at the source. Using the initial momentum and magnetic field, we calculate the polarization in the emitter frame and the Penrose-Walker constant at the source. 
\item Subsection~\ref{subsec:Redshift}: Using the conserved Penrose-Walker constant and the photon's arrival position, we calculate its polarization on the observer screen, taking redshift and path length effects into account.
\end{enumerate}

Subsection~\ref{subsec:FaceOn} then reviews simplifying aspects of the calculation in the special case of an on-axis observer.

\subsection{Light Propagation in Kerr} \label{subsec:LightProp}

We review photon propagation and polarization in Kerr following \cite{Himwich2020}. From the geodesic equation, a photon's (energy-rescaled) four-momentum is given in terms of its position and conserved quantities ($\lambda$,$\eta$) corresponding to the energy-rescaled angular momentum parallel to the axis of symmetry and Carter integral, respectively:
\begin{equation} \label{eq:fourmom}
p_{\mu}dx^{\mu} = - dt \pm_r \frac{\sqrt{\mathcal{R}(r)}}{\Delta(r)} dr \pm_{\theta} \sqrt{\Theta(\theta)}d\theta + \lambda d \phi,
\end{equation}
given in terms of the radial and angular potentials 
\begin{equation} 
  \begin{aligned}
	\mathcal{R}(r)&= (r^2+a^2-a\lambda)^2-\Delta\left[\eta+(a-\lambda)^2\right],\\
	\Theta(\theta)&=\eta+a^2\cos^2{\theta}-\lambda^2\cot^2{\theta}. \\
  \end{aligned}
\end{equation}
The photon trajectory is determined by its initial position, conserved quantities $(\lambda, \eta)$, and signs $\pm_r, \pm_{\theta}$ of its initial motion. The photon's linear polarization $f^\mu$ is parallel transported along its trajectory,
\begin{equation}
f^\mu p_\mu = 0, \ \ \ p^{\mu}\nabla_{\mu}f^{\nu} = 0.
\end{equation}
From the photon momentum and polarization, one can construct the Penrose-Walker constant $\kappa$, a complex constant conserved along the photon trajectory \cite{Walker_Penrose_1970,Chandrasekhar:1985}: 
\begin{equation} \label{eq:PW}
\begin{aligned}
  \kappa & = \kappa_1 + i \kappa_2 =  (\mathcal{A} - i\mathcal{B})(r - ia\cos\theta), \\
  \mathcal{A} &= (p^t f^r - p^r f^t ) + a \sin^2 \theta (p^r f^{\phi} - p^{\phi} f^r ),\\
    \mathcal{B} &= \left[ (r^2 + a^2) (p^{\phi} f^{\theta} - p^{\theta} f^{\phi}) - a  (p^t f^{\theta} - p^{\theta} f^t )\right]\sin\theta.
  \end{aligned}
\end{equation}
Given the Penrose-Walker constant, one can solve for the polarization $f^{\mu}$ at any point along the photon path. 

\subsection{Screen Coordinates \& Conserved Quantities} \label{subsec:BardeenCoords}

For a photon with conserved quantities ($\lambda$,$\eta$), its arrival position is given by the screen coordinates \cite{Cunningham1973} 
\begin{equation} \label{eq:Bardeen}
\alpha = - \frac{\lambda}{\sin \theta_o}, \ \ \ \beta = \pm_o \sqrt{\Theta(\theta)},
\end{equation}
where $\theta_o$ is the observer's polar inclination from the spin axis and $\pm_o$ is the sign of $p^{\theta}$ at the observer. 

Conversely, the photon's arrival position on the screen determines its corresponding conserved quantities:
\begin{equation} \label{eq:Bardeeninverse}
\begin{aligned}
\lambda &= - \alpha \sin\theta_o ,\\
\eta &= (\alpha^2 - a^2)\cos^2 \theta_o + \beta^2. \\
\end{aligned}
\end{equation}
For time-averaged, axisymmetric images of an equatorial disk, the relevant photon motion is in $(r,\theta)$. Photon trajectories from an initial position $(r_{\rm s}, \theta_{\rm s} = \frac{\pi}{2})$ to a final position $(r_o \rightarrow \infty, \theta_o)$ are given by the null geodesic equation \cite{Carter_1968}:
\begin{equation}
\label{eq:geoeq}
I_r = \fint_{r_{\rm s}}^{r_o} \frac{dr}{\pm_r\sqrt{\mathcal{R}(r)}} = \fint_{\theta_{\rm s}}^{\theta_o} \frac{d\theta}{\pm_r\sqrt{\Theta(\theta)}} = G_{\theta},
\end{equation}
where the slash denotes a monotonic path integral with the signs $\pm_r,\pm_{\theta}$ changing at radial and angular turning points, respectively. These have closed-form solutions in terms of elliptic integrals, which have been described in a variety of formalisms by many authors, including \cite{Cunningham_1975redshift,Rauch:1994,Agol:1997,Kapec2020}. We follow the conventions of \cite{Gralla2020b} (see references therein and App.~\ref{app:Definitions} for definitions). Following (81) of \cite{Gralla2020a}, for a trajectory with $m$ turning points in $\theta$ and $\theta_{\rm s} = \frac{\pi}{2}$ the geodesic equation becomes\footnote{
When $a=0$ exactly (for which (\ref{eq:geoeq}) is not well-defined) the geodesic equation reduces to (see e.g. Eq. B1 of \cite{Gates2020}):
\begin{align}
    I_r&=G_\theta=\frac{1}{\sqrt{\eta+\lambda^2}}\left[m\pi-{\rm sign}(\beta)\arcsin\left(\sqrt{1+\frac{\lambda^2}{\eta}}\cos\theta_o\right)\right].
\end{align} 
}
\begin{equation} \label{eq:intgeo}
I_{r} = G_\theta^m = \frac{1}{\sqrt{-u_{-}a^2}} \left( 2mK\left(\frac{u_+}{u_-}\right) - \text{sign}(\beta)F_o\right),
\end{equation}
where
\begin{equation} \label{eq:Fo}
F_o = F\left(\arcsin\frac{\cos{\theta_o}}{\sqrt{u_+}}\Big| \frac{u_+}{u_-}\right), 
\end{equation}
and 
\begin{equation}
u_{\pm} = \Delta_{\theta} \pm \sqrt{\Delta_{\theta}^2 + \frac{\eta}{a^2}}, \ \ \ \Delta_{\theta} = \frac{1}{2}\left(1 - \frac{\eta + \lambda^2}{a^2}\right).
\end{equation}
Given $m$ and $r_{\rm s}$, \eqref{eq:intgeo} defines a relationship between $\lambda$ and $\eta$ and therefore between image $\alpha$ and $\beta$ via \eqref{eq:Bardeen}.
\footnote{For $\theta_o>\pi/2$, (82) of \cite{Gralla2020a} generalizes to $\overline{m}=m-H(\beta\cos\theta_o)$. If $\theta_o<\pi/2$, the observer is above the midplane, so geodesics with odd $m$ arrive on the top half of the image, while geodesics with even $m$ arrive on the bottom. For $\theta_o>\pi/2$, the observer is below the midplane, so the parity of $m$ switches. See Fig. 7 of \cite{Gralla2020a}.}

The geodesic equation \eqref{eq:intgeo} can be inverted (see e.g. (30) of \cite{Gralla2020a}) to find $r_{\rm s}(I_r = G_{\theta}^m)$, which gives
\begin{equation} \label{eq:inversion}
r_{\rm s}(I_r) = \frac{r_4r_{31} - r_3r_{41} \text{sn}^2\left(\frac{1}{2}\sqrt{r_{31}r_{42}} I_r - \mathcal{F}_o \big| k\right)}{r_{31} - r_{41} \text{sn}^2\left(\frac{1}{2}\sqrt{r_{31}r_{42}} I_r - \mathcal{F}_o \big| k\right)},
\end{equation}
where
\begin{equation}
\mathcal{F}_o = F\left(\arcsin{\sqrt{\frac{r_{31}}{r_{41}}}} \Big| k \right),
\end{equation}
and
\begin{equation}
  k = \frac{r_{32}r_{41}}{r_{31}r_{42}}, \ \ \ r_{ij} = r_i - r_j, 
\end{equation}
with the roots $\left\{ r_i\right\}$ of $\mathcal{R}(r)$ (given below in App.~\ref{app:RadialRoots} as well as App.~A of \cite{Gralla2020a}). Because \eqref{eq:intgeo} defines a relation for $ r_{\rm s}(G_{\theta}^m)$, and $G_{\theta}^m$ can be written in terms of $(\lambda, \eta)$ or equivalently $(\alpha, \beta)$, \eqref{eq:inversion} defines an equation $r_{\rm s}(\alpha, \beta)$, which can be solved numerically to find the allowed curves of $(\alpha, \beta)$ that describe a source at radius $r_{\rm s}$. 

In practice, \eqref{eq:inversion} is computed using specified values of $a, \sin{\theta_o}, r_{\rm s}, \varphi\equiv\arctan(\beta/\alpha)$ and a test value(s) of $b\equiv \sqrt{\alpha^2+\beta^2}$ to solve for the corresponding impact parameter using {\fontfamily{qhv}\selectfont \small FindRoot} in \textit{Mathematica} 12 or {\fontfamily{cmtt}\selectfont
scipy.optimize} in {\fontfamily{cmtt}\selectfont
python 3}.\footnote{Notebooks available upon request to the corresponding author.} Here and throughout, $\varphi$ is the phase of $\alpha+i\beta$. Results of this calculation are displayed for a variety of $a$ and $\theta_o$ in Fig. 6 of \cite{Gralla2020a}.  

After computing the set of allowed screen positions $(\alpha, \beta)$ for a photon emitted at a given equatorial radius $r_{\rm s}$, \eqref{eq:Bardeeninverse} is used to determine the corresponding conserved quantities $(\lambda, \eta)$, which specify the photon momentum at the source and can be used to determine its polarization, as described in the next section.  

\subsection{Polarization in the Local Frame} \label{subsec:PolLocalFrame}

Having determined $(\lambda,\eta)$ for $r_{\rm s}, a, \theta_o$ as described in the previous section, the photon (energy-rescaled) momentum at the source is given by \eqref{eq:fourmom}:
\begin{equation}
  \begin{aligned}
    p_{t} &= - 1, \\
    p_{r} &=  \pm_r, \frac{\sqrt{\mathcal{R}(r_{\rm s})}}{\Delta_{\rm s}}, \\
    p_{\phi} &= \lambda, \\
    p_{\theta} &=  \pm_s \sqrt{\eta},  \\
  \end{aligned}
\end{equation}
where the sign of $p^{\theta}$ at the source is $\pm_s = (-1)^m \pm_o$.\footnote{For $\theta_o < \frac{\pi}{2}$, $\pm_s = -1$ for both $m =0$ on the bottom of the image (where $\pm_o = -1$) and $m = 1$ on the top (where $\pm_o = 1$). \label{foot:m}} 

The sign $\pm_r$ depends on $\{\lambda,\eta, r_{\rm s}, a, \theta_o,m\}$ and must be computed semi-analytically as described in Appendix \ref{app:Rmotion}. With the sign of $p^r$ determined, we compute the components of the four-momenta at the source as 
\begin{equation} \label{eq:upperp}
  \begin{aligned}
    p^t&=\frac{1}{r_{\rm s}^2}\left(-a(a - \lambda)+\frac{(r_{\rm s}^2+a^2)(r_{\rm s}^2 + a^2 - a \lambda)}{\Delta_{\rm s}} \right),\\
	p^r&= \pm_r \frac{1}{r_{\rm s}^2}\sqrt{\mathcal{R}(r_{\rm s})}, \\ 
	p^{\phi}&=\frac{1}{r_{\rm s}^2} \left(- (a - \lambda) + \frac{a}{\Delta_{\rm s}}\left(r_{\rm s}^2+a^2 - a \lambda\right)\right),\\
	p^{\theta}&= \pm_s \frac{\sqrt{\eta}}{r_{\rm s}^2}. \\
\end{aligned}
\end{equation}
To compute the local photon polarization at the source, $p^{\mu}$ is transformed to the local frame of the emitter via \eqref{eq:boostedframe}. In the local frame $f^{(t)} = 0$, and the spatial components $\vec{f} = \left(f^{(r)},  f^{(\phi)}, f^{(\theta)}\right)$ are given by a cross product of the local three-momentum $\vec{p} = \left(p^{(r)},  p^{(\phi)}, p^{(\theta)}\right)$ with the local magnetic field  $\vec{B} = \left(B^{(r)},  B^{(\phi)}, B^{(\theta)}\right)$:
\begin{equation} \label{eq:polcross}
  \vec{f} = \frac{\vec{p} \times \vec{B}}{|\vec{p}|},
\end{equation}
as expected for synchrotron radiation \cite{Rybicki_1979,Narayan_2020}. Note that the axes $(\hat{r}), (\hat{\phi}),$ and $(\hat{\theta})$ are orthogonal in the fluid frame with $(\hat{r})\times (\hat{\phi})=(\hat{\theta})$, so the cross product has its standard form in $\mathbb{R}^3$. For later reference, we record explicitly:
\begin{equation} \label{eq:crossprod}
\begin{aligned}
f^{(r)} &\propto p^{(\phi)}B^{(\theta)} - p^{(\theta)}B^{(\phi)},\\
f^{(\phi)} &\propto - p^{(r)}B^{(\theta)} + p^{(\theta)}B^{(r)},\\
f^{(\theta)} &\propto  p^{(r)}B^{(\phi)} - p^{(\phi)}B^{(r)}.\\
\end{aligned}
\end{equation}

The intensity of synchrotron radiation emitted along $\vec{p}$ depends on $\sin\zeta$, where $\zeta$ is the pitch angle between $\vec{p}$ and $\vec{B}$ \citep{Narayan_2020}:
\begin{equation} \label{eq:sinzeta}
\sin \zeta = \frac{| \vec{p} \times \vec{B}|}{|\vec{p}||\vec{B}|}. 
\end{equation}
By construction, the intensity has magnitude 
\begin{equation}
f^{\mu}f_{\mu} = \sin^2\zeta |\vec{B}|^2.
\end{equation}

We model idealized axisymmetric pointlike emission, following \cite{Narayan_2020}.  Our model therefore predicts relative values of the polarized intensity across an image varying only due to non-constant pitch angle and redshift factor, and does not capture the variations present due to the electron distribution function and changing plasma properties. Our model does not contain an absolute scale for the polarized intensity, as we later take $|\vec{B}|=1$. Moreover, we do not consider the partial incoherence of synchrotron radiation, and we treat factors that contribute to the emissivity as constant (such as variations in the plasma density or temperature). This effectively predicts the polarized intensities up to proportionality constants. Since no polarized radiative transfer occurs outside of the pointlike emitters, we further disregard all Faraday effects, setting the polarization fraction to unity.

Given $f^{(a)}$ in the local frame, $f^{\mu}$ in Boyer-Lindquist coordinates is computed via \eqref{eq:invtetrad}. Then, the Penrose-Walker constant \eqref{eq:PW} for the trajectory is computed from $f^{\mu}$ and $p^{\mu}$ at the source $r_{\rm s}, \theta_{\rm s} = \frac{\pi}{2}$ (see \eqref{eq:upperp}), which simplifies to 
\begin{equation}
\label{eq:kappadef}
  \begin{aligned}
  \kappa &= \kappa_1 + i \kappa_2 = r_{\rm s}\left(\mathcal{A} - i \mathcal{B} \right),\\
    \mathcal{A} &= (p^t f^r - p^r f^t) + a (p^r f^{\phi} - p^{\phi} f^r),\\
    \mathcal{B} &= \left[ (r_{\rm s}^2 + a^2) (p^{\phi} f^{\theta} - p^{\theta} f^{\phi}) - a  (p^t f^{\theta} - p^{\theta}f^t) \right], 
  \end{aligned}
\end{equation}
giving the simple relation
\begin{equation}
\left(\kappa_1, \kappa_2 \right) = r_{\rm s} \left(\mathcal{A}, - \mathcal{B} \right)
\end{equation}
for equatorial sources.\footnote{An analogous calculation to analytically compute geodesics and parallel transport $f$ is implemented in the ray-tracing code \texttt{grtrans} \cite{Dexter_2016}. Our model differs from this code because we ray-trace \emph{forwards} from a fixed emission radius, as opposed to backwards from a fixed impact parameter. Furthermore, we use the Legendre elliptic formalism presented by \cite{Gralla2020a} as opposed to the Carlson elliptic formalism presented by \cite{Dexter_2009}, and we formulate the fluid-frame tetrad entirely in terms of Lorentz boosts as opposed to using the results of a Gram-Schmidt orthogonalization.}

\subsection{Observed Polarization and Redshift} \label{subsec:Redshift}

Given the photon arrival position $(\alpha, \beta)$ and its Penrose-Walker constant $\kappa$, computed as described in the previous subsections, the observed polarization (direction of electric field transverse to photon momentum) is computed from the components of $f^{\mu}$ at large radius projected along the $\hat{\alpha}$ and $\hat{\beta}$ directions on the observer screen \cite{Himwich2020}:
\begin{equation} \label{eq:screenpol}
\begin{aligned}
\left(f^{\alpha},f^{\beta}\right) &= \frac{1}{\mu^2 + \beta^2}\left(\beta \kappa_2 - \mu \kappa_1, \beta \kappa_1 + \mu \kappa_2\right), \\
\mu &= - (\alpha + a \sin \theta_o).
\end{aligned}
\end{equation}
Here, ($f^\alpha$,$f^\beta)$ is a two-vector on the asymptotic observer screen (see Appendix A of \cite{Himwich2020} for details).

The photon conserved energy $E =1$ is the energy measured by stationary observers at infinity. The energy measured in the rest frame of the emitting source is
\begin{equation}
  E_s = p^{(t)} = - p_{\mu}u^{\mu}, 
\end{equation}
and the ratio of these two energies is the redshift,
\begin{equation}
  g = \frac{E}{E_s} = \frac{1}{p^{(t)}}.
\end{equation}
Since $I_\nu/\nu^3$ is invariant along a geodesic, the specific intensity is Doppler boosted by a factor of $g^3$ when it reaches the observer.

The emitted intensity varies with frequency $\nu$  as $I_{\nu} \sim \nu^{-\alpha_{\nu}}$, with angular dependence $\left(\sin \zeta\right)^{1 + \alpha_{\nu}}$. A nonzero spectral index in turn causes specific intensity to be boosted by an additional factor of $g^{\alpha_\nu}$ when it reaches the observer,\footnote{For a spectral index $\alpha_\nu$, the invariance of $I_\nu/\nu^3$ implies that $I_{\nu,o}/\nu_{o}^3=I_{\nu,s}/\nu_s^3=\nu_s^{-(3+\alpha_\nu)}$, where $o$ denotes observer and $s$ denotes source. Since $\nu\propto p^{(t)}$, one has $I_{\nu,o}=\nu_o^3/\nu_s^{3+\alpha_\nu}=E_s^{-(3+\alpha_\nu)}=g^{3+\alpha_\nu}$.} giving $I_{\nu, o}=I_{\nu,s}g^{3+\alpha_\nu}$. Note that in this paper, ``intensity" (``flux") always refers to \textit{specific} intensity (flux). For an optically and geometrically thin disk, which we assume throughout Section~\ref{sec:Visualizations}, intensity grows linearly with the geodesic path length $l_p$ through the emitting material \cite{Narayan_2020}: \begin{align}
    \l_p&=\frac{p^{(t)}_s}{p^{(z)}_s}H,
\end{align}
with $H$ the height of the disk, taken to be a constant. Since the intensity is proportional to the square of the electric field (i.e. the square of the polarization vector), the observed components of the polarization vector are proportional to the square root of the path length $l_p$ and boost of $g^{3+\alpha_\nu}$:
\begin{equation} \label{eq:PolObserved}
    \left(f^{\alpha}_{\text{obs}}, f^{\beta}_{\text{obs}}\right) \propto \sqrt{l_p} g^{\frac{3+\alpha_\nu}{2}}\left(f^{\alpha}, f^{\beta}\right).
\end{equation}
Note that following \cite{Himwich2020}, we measure the Electric Vector Position Angle (EVPA) counter-clockwise from $+\hat{\beta}$: \begin{align} \label{eq:EVPA}
    \EVPA\equiv \arctan\left({-\frac{f^\alpha_{\rm obs}}{f^\beta_{\rm obs}}}\right).
\end{align}
Our model assumes pure synchrotron radiation with a constant polarization fraction of 1. Additionally, in this paper, we take $\alpha_{\nu} \sim 1$, which is consistent with M87* observed at 230 GHz \cite{Narayan_2020} and with the range of values $\alpha_{\nu} \sim 0.5-1.5$ that describe very bright flares ($\alpha_{\nu} \sim 0.5$) to the average lower-flux density state ($\alpha_{\nu} \sim 1.5$) of Sgr~A* (see \cite{Witzel:2018kzq} and references therein).

\subsection{Face-On Coordinates and EVPA} \label{subsec:FaceOn}

This subsection reviews the special case of an on-axis observer. For a black hole viewed face-on $(\theta_o=0^\circ)$, the coordinates $(\alpha,\beta)$ on the screen degenerate because the $\hat{\beta}$ axis becomes a point. Still, the radial screen coordinate
\begin{equation}
  b = \sqrt{\alpha^2 + \beta^2} = \sqrt{\eta + \lambda^2 + a^2\cos^2\theta_0} \rightarrow \sqrt{\eta + a^2}
\end{equation}
remains well-defined. There is a single spherical photon orbit radius that crosses the spin axis, corresponding to photons with $\lambda = 0$. Taking $(b, \varphi = \arctan \frac{\beta}{\alpha})$ in the limit $\lambda \to 0$ $(\theta_o \to 0)$ defines Cartesian coordinates $(b \cos \varphi, b \sin \varphi)$ that smoothly interpolate to the usual definition for observers with $\theta_o > 0$. The polarization components are
\begin{equation} 
    \label{eq:ffaceon}
    (f^\alpha,f^\beta) \propto \left(b\sin\varphi\kappa_2+b\cos\varphi\kappa_1, b\sin\varphi\kappa_1-b\cos\varphi\kappa_2\right),
\end{equation}
and the EVPA is given by 
\begin{equation} \label{eq:FaceOnEVPA}
    \EVPA = \arctan\left(-\frac{\kappa_2\sin\varphi+\kappa_1\cos\varphi}{\kappa_1\sin\varphi-\kappa_2\cos\varphi}\right).
\end{equation}
The direct image of an axisymmetric source for a face-on observer is entirely described by $m=0$ with $\pm_s=-1$, and the geodesic equation for all spins is approximated well by the simple relation of ``just adding 1,'' $b \approx r_{\rm s} + 1$ ($M =1$) \cite{Gralla2020a}. Including subleading terms  (see Eq.~10 of \cite{Gates2020}),
\begin{align}
    r_{\rm s} &= b-1 + \frac{1-a^2}{2b} + \frac{3(5\pi-16)}{4b^2} + \mathcal{O}(1/b^3),
\end{align}
which may be inverted to yield
\begin{align} \label{eq:bseries}
    b &= r_{\rm s}+1+\frac{a^2-1}{2r_{\rm s}} + \frac{50-2a^2-15\pi}{4r_{\rm s}^2} + \mathcal{O}(1/r_{\rm s}^3).
\end{align}
We will make use of this analytic expansion to quantify the effects of spin on observed EVPA in Section \ref{sec:EVPA}.

\section{Polarization Visualizations} \label{sec:Visualizations}

\begin{figure*}[t]
    \centering
    \includegraphics[width=0.99\textwidth]{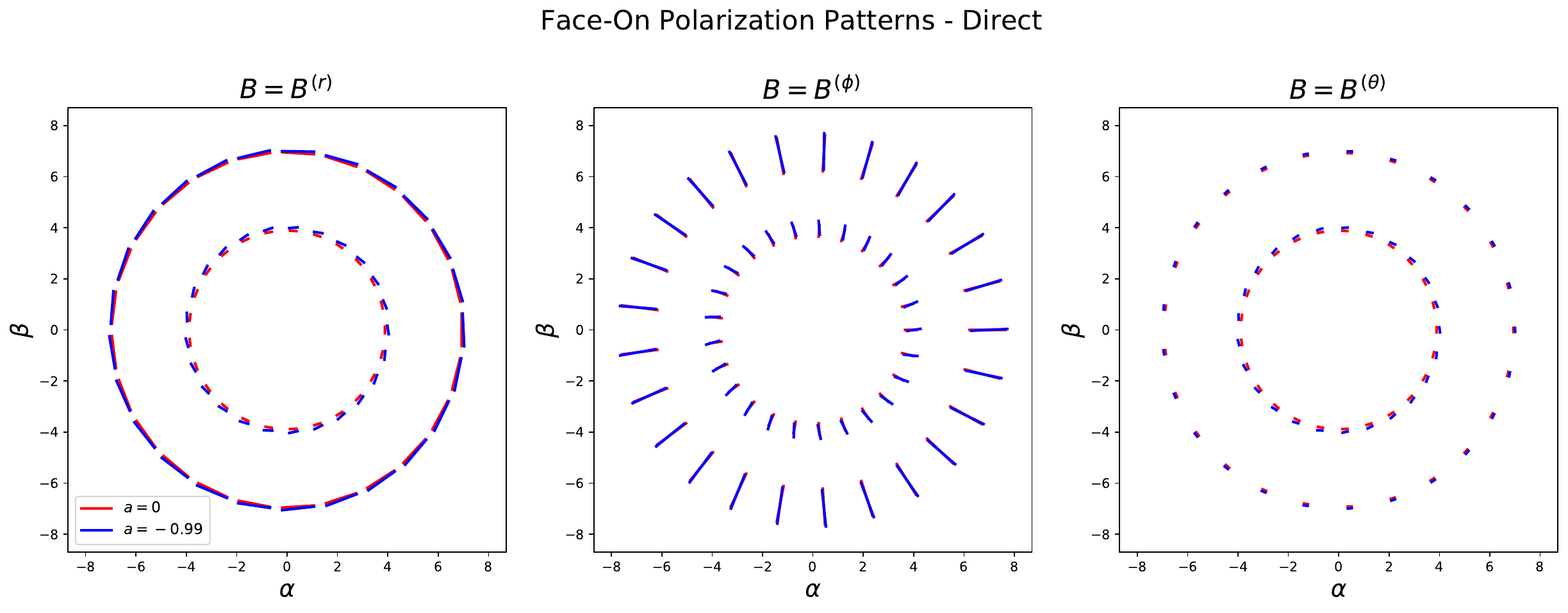}  
    \caption{Polarized intensity tick plots for three idealized magnetic field configurations: from left to right, radial $B = B^{(r)}$, azimuthal $B = B^{(\phi)}$, and vertical $B = B^{(\theta)}$, in the case the direct image seen by an on-axis observer $\theta_o=0$. The fluid is modelled by an unboosted ZAMO emitter ($\beta_v=0$ in \eqref{eq:betaboost}). Each plot shows two spins ($a=0$ and $a=-0.99$ in red and blue, respectively), as well as two emission radii ($r_{\rm s}=3$ and $r_{\rm s}=6$, corresponding to the inner and outer rings, respectively). 
    } 
    \label{fig:faceonticks}
\end{figure*}
\begin{figure*}[t]
    \centering
    \includegraphics[width=\textwidth]{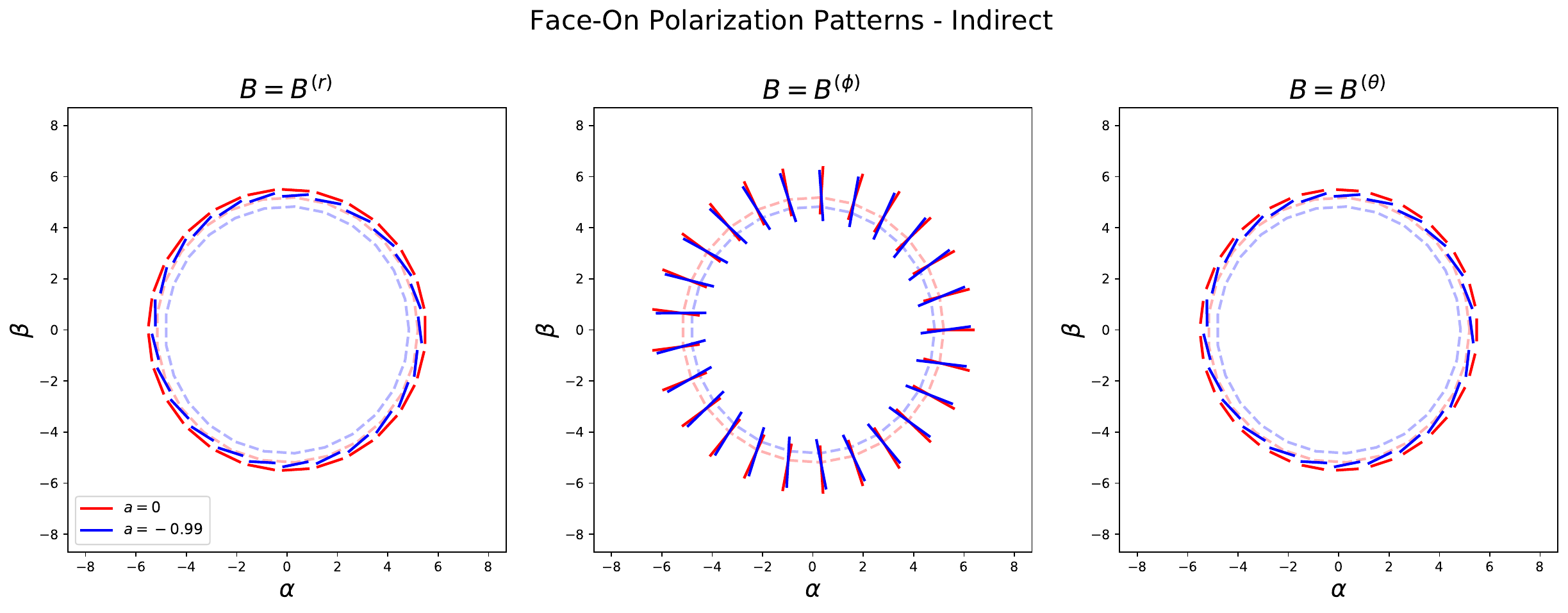}
    \caption{Polarized intensity tick plots for the indirect image corresponding to the direct image shown in Fig.~\ref{fig:faceonticks} for  $\theta_o=0$ and $\beta_v=0$, now displayed for the single emission radius $r_{\rm s}=6$, and with critical impact parameters shown as dashed lines. 
    }
    \label{fig:faceonsubimageticks}
\end{figure*}

In this section, we use our model to compute the polarized image of equatorial emitting sources at a variety of different spins, magnetic fields, and inclination angles. Using \eqref{eq:PolObserved}, we compute the polarized intensity of various configurations and present the resulting images as ``tick plots" as in \cite{Narayan_2020}. To draw comparisons to M87*, we consider $a < 0$, $\theta_o < \frac{\pi}{2}$, corresponding to the clockwise fluid motion on the sky seen around M87*.\footnote{Note that this terminology differs from that used in EHT Paper V \cite{PaperV}, in which $a<0$ corresponds to retrograde accretion flow (counterclockwise on the sky). For us, $a<0$ corresponds to a prograde accretion flow (clockwise on the sky) with the spin axis pointed away from the observer.} In each tick plot we display $a=-0.99$ to represent the extremal Kerr limit, as exact expressions for $|a|=1$ and $\theta_o = \frac{\pi}{2}$ require additional care (see e.g. \cite{Gates2020}). The relationship between positive and negative spin is discussed in detail in Appendix~\ref{app:symmetries}.

\subsubsection*{On-axis Observers}

Three sample tick plots for on-axis observers of purely radial, azimuthal, and vertical magnetic fields are presented respectively in the left, middle, and right panels of Figure~\ref{fig:faceonticks}. We emphasize that these purely radial, azimuthal, and vertical magnetic field geometries are idealized and are specified pointwise. They have been chosen to probe the breadth of potential magnetic field geometries that could be present in realistic horizon-scale black hole accretion flows or subsections of accretion flows. In reality, physical magnetic fields vary throughout the accretion region and contain nonzero radial, azimuthal, and vertical components. Each tick has orientation aligned with the EVPA of the arriving photon and length proportional to intensity (square of the electric field). In each panel, the inner and outer circles correspond to emission radii of $r_{\rm s}=3$ and $r_{\rm s}=6$, respectively, and the boost parameter from \eqref{eq:betaboost} is taken to be $\vec{\beta}_v=0$. As expected from the axisymmetric model viewed face-on, all three panels of Figure~\ref{fig:faceonticks} are rotationally symmetric. The ticks in the rightmost panel, which displays a vertical magnetic field, are significantly shorter than those in the cases of radial and azimuthal fields; for purely vertical fields $\vec{B}=B^{(\theta)}$ in \eqref{eq:sinzeta}, $|\sin\zeta|\propto\vec{p}\times \vec{B}\ll 1$ and is only nonzero due to relativistic effects \cite{Gravity_2020}.

In addition to the direct image, our model includes general subimages, also referred to as indirect images, which are indexed by number of radial turning points $m$. For large $m$ in the face-on case, the impact parameter $b$ will approach the critical impact parameter $b_c$, up to exponentially suppressed corrections \cite{Gralla2020a,Johnson_2020}. The definition of critical parameters and the critical curve corresponding to bound photon orbits are presented in Appendix~\ref{app:criticalcurve}. For illustration, tick plots for the first subimage corresponding to the direct image in Figure~\ref{fig:faceonticks} are shown in Figure~\ref{fig:faceonsubimageticks}, now displayed for clarity for just a single emission radius $r_{\rm s}=6$. The critical impact parameters $b_c$  are also shown as dashed lines. Note that all of the tick plots display relative polarized \textit{intensity} from \eqref{eq:PolObserved}. The polarized flux from subimages is suppressed by a demagnification factor, which is described in  Appendix~\ref{app:magnification}. When direct and indirect images are summed, this can cause depolarization in the total image \cite{Jimenez-Rosales:2021ytz}.

The low-spin and high-spin signatures in Figure~\ref{fig:faceonticks} showcase relatively similar EVPAs. The effect of spin becomes slightly more visible at smaller emission radii due to the stronger lensing  effects of the higher curvature nearer to the black hole. Relative to Figure~\ref{fig:faceonticks}, the photon arrival positions in Figure~\ref{fig:faceonsubimageticks} have all been pushed closer to the critical curve, and the distinction between the low-spin and high-spin EVPAs is consequently more pronounced in the indirect image than in the direct image. We quantify these effects for the face-on case in Section \ref{sec:EVPA}.

\begin{figure*}[t]
    \centering
    \includegraphics[width=\textwidth]{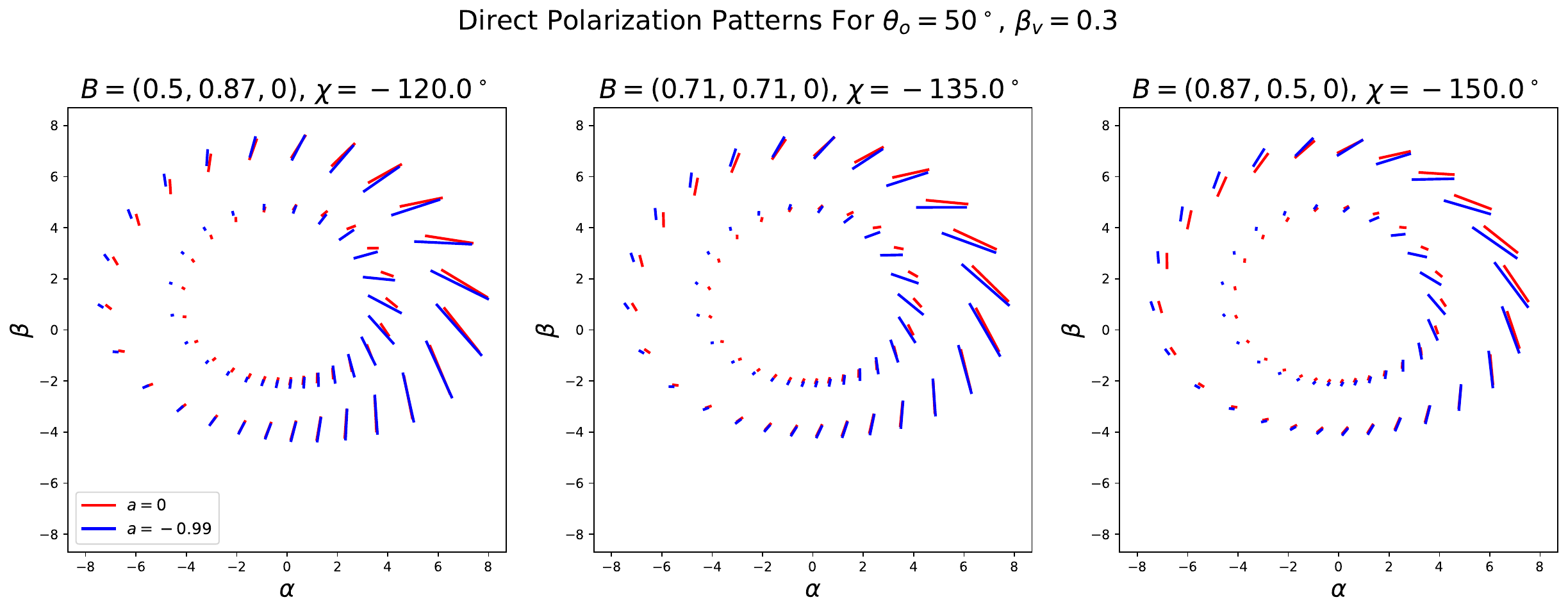}
    \caption{Polarized intensity tick plots for direct image with $\theta_o=50^\circ$ and boost parameter $\beta_v=0.3$, shown for various equatorial field configurations $\vec{B} = (B^{(r)}, B^{(\phi)}, B^{(\theta)})$ and fluid velocities $\chi$ (see \eqref{eq:betaboost}). The fluid configurations are taken, from left to right, to match those in the top left, top right, and bottom left of Figure 5 of \cite{Narayan_2020}, with boost direction anti-parallel to the magnetic field. Each plot shows two spins ($a=0,-0.99$ in red and blue, respectively) and two emission radii ($r_{\rm s}=3,6$, the inner and outer rings, respectively).
    }
    \label{fig:highincticksdirect}
\end{figure*}
\begin{figure*}[t]
    \centering
    \includegraphics[width=\textwidth]{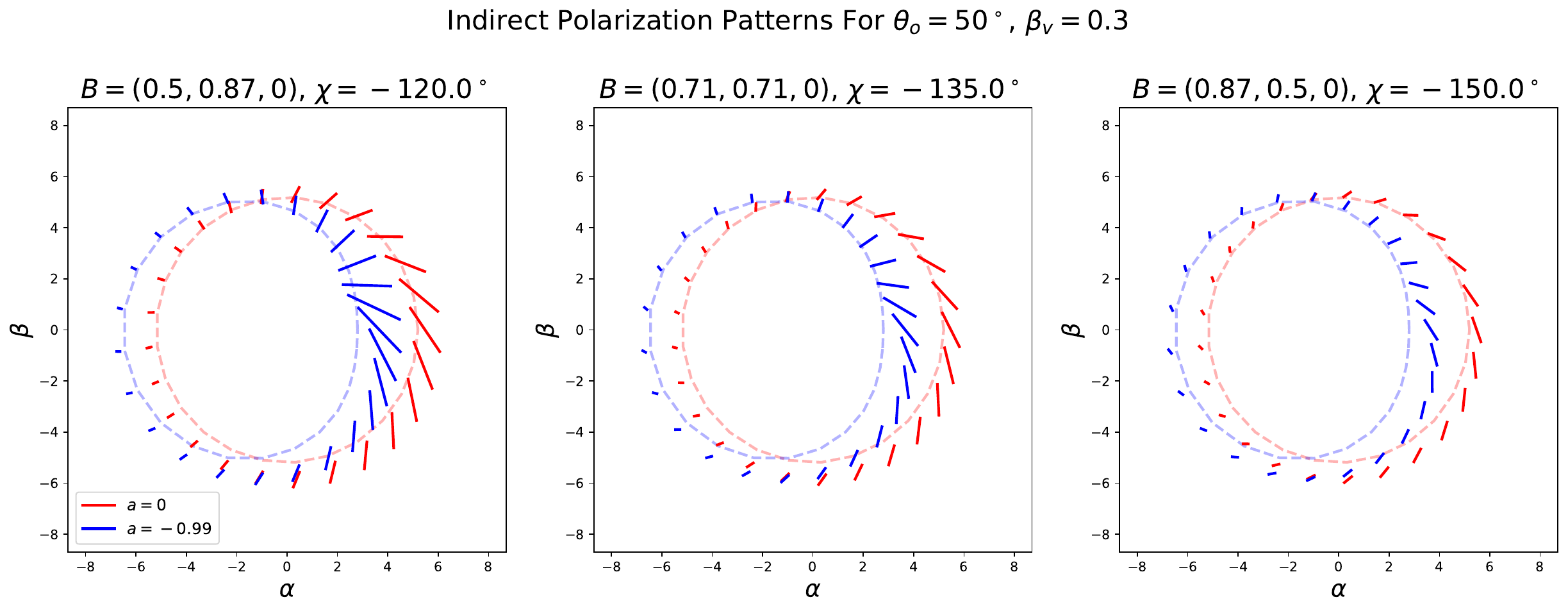}
    \caption{Polarized intensity tick plots for the indirect images corresponding to the direct images shown in Fig.~\ref{fig:highincticksdirect} for  $\theta_o=50^\circ$ and $\beta_v=0.3$, displayed for the single emission radius $r_{\rm s}=6$ with critical impact parameters shown as dashed lines. 
    }
    \label{fig:highincticksindirect}
\end{figure*}

\subsubsection*{Inclined Observers}
The effects of spin on the observed polarization pattern become more pronounced with increasing observer inclination $\theta_o$, as one may intuitively expect from the appearance of $a\sin\theta_o$ in \eqref{eq:screenpol}.  An illustrative example is shown in Figure~\ref{fig:highincticksdirect}, which displays tick plots for the direct image of various equatorial field configurations $\vec{B} = (B^{(r)}, B^{(\phi)}, B^{(\theta)})$ and boost directions $\chi$ (see \eqref{eq:betaboost}) observed at a moderately high inclination $(\theta_o=50^\circ)$ and boost parameter ($\beta_v=0.3$). Again, the inner and outer circles correspond to emission radii of $r_{\rm s}=3$ and $r_{\rm s}=6$, respectively. The fluid rotation and boost configurations are taken, from left to right, to match those in the top left, top right, and bottom left of Figure 5 of \cite{Narayan_2020}. The boost direction is chosen so that it is parallel to the magnetic field. As in the face-on case, the effect of spin becomes stronger at smaller emission radii. Due to the moderately high observer inclination, in all three panels, the bottom half of the image (corresponding to emission from the near half of the midplane) is pushed towards the $\alpha$ axis and has significantly lower intensity. 

Note that for the cases of purely radial and toroidal fields (left two panels) the polarized intensity at $r_{\rm s} = 6$ is greater than at $r_{\rm s} = 3$, while the reverse holds for vertical fields (right panel). This is due to the stronger lensing effects at smaller radii, which increase the relative local $p^{(r)}$ and therefore decrease the pitch angle \eqref{eq:sinzeta} for $B = B^{(r)}, B^{(\phi)}$ and increase it for $B = B^{(\theta)}$ (see \eqref{eq:crossprod}).

In Figure~\ref{fig:highincticksindirect}, we also show the corresponding tick plots for the first subimage ($m=1$ for bottom half of image and $m=2$ for top half of image), now displayed for clarity for only a single emission radius $r_{\rm s}=6$. As in Figure~\ref{fig:faceonsubimageticks}, we overlay the critical curves for the low-spin and high-spin geometries as dashed lines. The high-spin critical curve is substantially displaced from the image origin, significantly affecting the photon arrival position. The corresponding  polarization is altered both by the change in arrival position and the explicit dependence of polarization on spin in \eqref{eq:screenpol}. Because the photon arrival positions are closer to the critical curve, the subimage shows a much higher degree of symmetry across the $\hat{\alpha}$ axis than the direct image. Note that compared with the direct image, the intensity in the indirect image is no longer diminished for vertical fields because photons in the indirect image experience stronger lensing effects and their emission angle \eqref{eq:sinzeta} relative to the vertical magnetic field no longer must be small. 

\section{EVPA of On-Axis Observers} \label{sec:EVPA}

This section presents an approximate analytic expression for the  effect of spin on EVPA in face-on images, and demonstrates the accuracy of this approximation using the toy model. 

\subsection{Direct Image} \label{subsec:DirectEVPA}

\begin{figure*}[t]
    \centering
    \includegraphics[width=\textwidth]{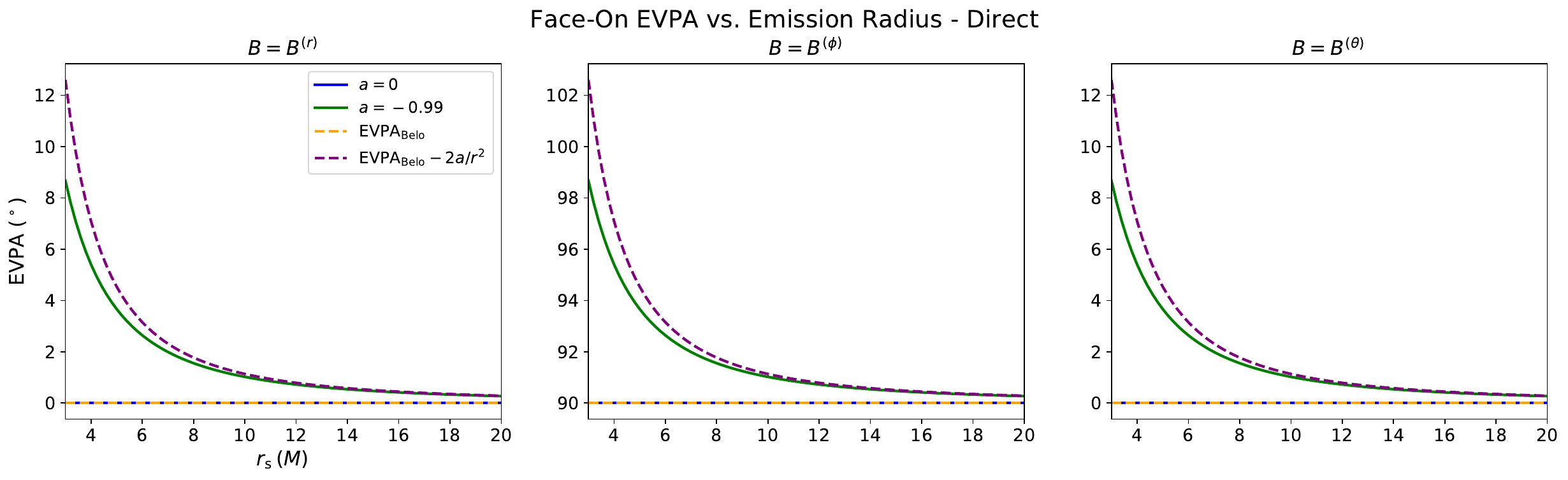}
    \caption{Face-on EVPA as a function of emission radius (in units of $M$) at $\varphi = 0$ on the image for purely radial $B = B^{(r)}$, toroidal $B = B^{(\phi)}$, and vertical $B = B^{(\theta)}$ magnetic fields. The numerically computed values are shown for $a =0$ and $a = -0.99$ in blue and green, respectively, and the approximations for zero-spin (Beloborodov) and high-spin in \eqref{eq:faceonapprox} are shown as dashed yellow and purple lines, respectively.     }
    \label{fig:faceoncomparedirect}
\end{figure*}

As described in Section~\ref{subsec:FaceOn}, the direct image seen by an on-axis observer corresponds to source momenta with $\lambda = 0$ and is parametrized by the impact parameter $b = \sqrt{\eta + a^2}$. To leading order, $b$ is related to the source radius $r_{\rm s}$ by ``just adding 1'' with subleading corrections given in \eqref{eq:bseries}. Computing $(\kappa_1,\kappa_2)$ as described in Section \ref{subsec:PolLocalFrame} with $(\lambda = 0, \eta = \sqrt{b^2 - a^2})$ yields an expression for the EVPA as a function of $b$ and $r_{\rm s}$, which may be series expanded in $r_{\rm s}$ using \eqref{eq:bseries}. Applying this expansion to \eqref{eq:FaceOnEVPA} with $\beta_v=0$, we obtain 
\begin{align}\label{eq:dchifaceon} 
    &\EVPA = 
    \nonumber\eta_e+\varphi+\frac{B^{(\theta)}\sin\eta_e}{B_{\rm eq}r} -\frac{2a}{r^2} \\
    \nonumber&\ \ +\frac{\sin\eta_e\left(16\cos\eta_e(B_{\rm eq}^2 {+} 2 (B^{(\theta)})^2) {+} 3 B_{\rm eq} B^{(\theta)}(5\pi {-} 16)\right)}{8B_{\rm eq}^2r^2} \\
    &\ \ +\mathcal{O}\left(\frac{1}{r_{\rm s}^3}\right)
\end{align}
with $\vec{B}_{\rm eq}\equiv B_{\rm eq}(\cos\eta_e(\hat{r})+\sin\eta_e(\hat{\phi}))$ (as in \cite{Narayan_2020}). The leading order correction to EVPA due to spin is $-2a/r^2$, which is independent of the magnetic field and hence a purely geometric effect. The effect of spin is suppressed by two orders of magnitude compared to the leading order term, explaining the similarity between the low-spin and high-spin EVPAs seen in Figure~\ref{fig:faceonticks}. 

As discussed in detail in \cite{Narayan_2020}, the Schwarzschild impact parameter is extremely well-approximated by the Beloborodov relation \cite{Beloborodov_2002}, which for a face-on black hole is (compare $b^2 = r_{\rm s}^2 + 2r_{\rm s} + a^2 +\mathcal{O}(r_{\rm s}^{-1})$ using \eqref{eq:bseries}):
\begin{align}
    b_{\rm Belo} = \sqrt{r_{\rm s}(2+r_{\rm s})}.
\end{align}
Defining $\EVPA_{\rm Belo}\equiv \EVPA(b_{\rm Belo})$, this implies\footnote{We also remark that we have not yet found an accurate series expansion akin to \eqref{eq:faceonapprox} for arbitrary inclinations.}
\begin{align}\label{eq:faceonapprox}
    \EVPA &\approx \EVPA_{\rm belo} -\frac{2a}{r_{\rm s}^2},
\end{align}
which holds in the limit where $B_{\rm eq}\to 0$, even though \eqref{eq:dchifaceon} will not be well-defined. One can also show that this relation holds for nonzero $\beta_v$, indicating that the primary effect of frame-dragging on polarization does not depend on the orbital speed of the accreting material. Note that the minus sign in  \eqref{eq:faceonapprox} implies that increasing $a$ will cause the EVPA to rotate in the opposite direction of the spin. 

The $-2a/r_{\rm s}^2$ correction matches the calculations performed by \cite{Pineault_1977,Fayos_1982,Brodutch_2011} and first termed  ``gravitational Faraday rotation" by \cite{Ishihara_1988}.\footnote{
The setup of our problem is slightly different than that of \cite{Pineault_1977,Fayos_1982,Brodutch_2011}. While we fix $\varphi$, they effectively fix $\phi$. The results that appear within the calculation, however, still match. If one instead fixes $\phi$, the arrival coordinate $\varphi$ will shift by $\sim 2a/r_{\rm s}^2$ and the net polarization rotation will be $\mathcal{O}(r_{\rm s}^{-3})$ \cite{Pineault_1977, Fayos_1982}. Weak-field expansions performed by \cite{Ishihara_1988,Nouri_1999} show a leading order correction of $\mathcal{O}(r_{\rm min}^{-3})$, with $r_{\rm min}$ the largest root of the radial potential.} Past work  derived this result by analyzing local reference frames/directions at every point on the geodesic in question. Because such reference frames are not unique, the ambiguity historically led to disagreements about the effects of spin on parallel transport \cite{Fayos_1982}. By combining conservation of $\kappa$ with the ``just add one" approximation, we entirely overstep this issue, providing a clean approximation of the gravitational Faraday effect that is recast in terms of only the emitter and observer frames.

In Figure~\ref{fig:faceoncomparedirect}, we test the accuracy of $\eqref{eq:faceonapprox}$ by plotting the EVPA as a function of emission radius (scaled in units of $M$) at $\varphi = 0$, on the image for purely radial $B = B^{(r)}$, toroidal $B = B^{(\phi)}$, and vertical $B = B^{(\theta)}$ magnetic fields. By axisymmetry, the EVPA at an arbitrary angle $\varphi = \vartheta$ is offset by exactly $\vartheta$ from the EVPA at $\varphi=0$. The numerically computed values are shown for $a = 0$ and $a = -0.99$ in blue and green, respectively, and the approximations for $a = 0$ (Beloborodov) and $a = 1$ in \eqref{eq:faceonapprox} are shown as dashed yellow and purple lines, respectively. The approximation \eqref{eq:faceonapprox} is demonstrated to hold reasonably well in all cases for $r\gtrsim 6M$. In the case of radial and toroidal fields with $\eta_e = 0$ and $\eta_e = \pi$, respectively, the EVPA will asymptote to $\eta_e + \varphi$ (up to additive factors of $\pi$). For the purely vertical field, the EVPA asymptotes to $\varphi$. 

Note that the Schwarzschild EVPA is constant over emission radius due to spherical symmetry of the Schwarzschild solution, which implies that fixed screen angle  $\varphi$ corresponds to fixed emission angle $\phi$ (see \eqref{eq:schphi} below). In this case, the EVPA at fixed $\varphi$ remains unchanged as a function of $r_{\rm s}$ for magnetic fields that are aligned with the local frame axes. 
 
Since the effect of spin is to rotate all polarization ticks by an equal amount $(-2a/r_{\rm s}^2)$, the geometric imprint of spin on polarization is indistinguishable from that of electromagnetic Faraday rotation by a coherent external screen at a fixed observing frequency. Therefore, single-frequency observations at the resolution of the EHT will likely be unable to determine black hole spin from the geometric rotation alone. However, because the geometric rotation is achromatic while Faraday effects are chromatic, it could be isolated by observations at multiple frequencies. In addition, if the magnetic field and spin of the black hole can be determined via other methods, deviations from the uniform geometric rotation could provide information about internal Faraday rotation, which causes differential relative rotation across the image.

While the geometric effect of spin on photon trajectories is subleading, spin can strongly influence the accretion dynamics and emissivity profile of the plasma surrounding the black hole. For instance, the location of the innermost stable circular orbit (ISCO) depends strongly on spin and in some cases may control where the accretion disk truncates, altering the resultant appearance of the black hole. This principle has helped guide numerous measurements of black hole spin using both X-ray reflection spectroscopy and continuum fitting \cite{Remillard:2006,Reynolds_2013x,Mcclintock_2011,Brenneman_2013}). Comparison of the simple geometric approximation \eqref{eq:faceonapprox} with observed data and GRMHD simulations could provide additional ways to distinguish geometric effects of spin from these astrophysical effects of spin. These phenomena, which cannot be disentangled in observed data, are also difficult to disentangle in GRMHD simulations. The GRMHD simulations are evolved from initial conditions in a fully general relativistic framework, allowing the effects of spin simultaneously change the geometry, the distribution of matter, and the magnetic field configurations therein \cite{PaperV}. 

It is instructive to compare the results of \eqref{eq:faceonapprox} to the work of \cite{Palumbo_2020}, which analyzed the EHT GRMHD library by examining rotational modes of images with low-inclination $(17^\circ)$ and found that the phase of the second rotational Fourier mode (which describes rotational symmetry and is denoted $\beta_2$ \cite{Palumbo_2020}) shifted by over $40^\circ$ from $a=0$ to $a=0.94$. By contrast, \eqref{eq:faceonapprox} suggests that for an emission radius of e.g. $r_{\rm s}=6$ (the Schwarzschild ISCO), $\beta_2$ should only rotate by $4a/r_{\rm s}^2\sim 6^\circ$. The spin-dependent effects present in the images analyzed by \cite{Palumbo_2020} are therefore likely a result of the spin-dependent accretion dynamics and not of the changing underlying geometry in which the geodesics propagate.

\subsection{Subimages}

The preceding face-on EVPA approximation may be generalized to subimages using the fact that for large $m$, the impact parameter $b$ approaches the critical impact parameter $b_c$ up to exponentially suppressed corrections \cite{Gralla2020a,Johnson_2020}. Performing the series expansion of \eqref{eq:FaceOnEVPA} in $r_{\rm s}^{-1}$ with the replacement $b \to b_c$, one finds that the frame dragging produces an $\mathcal{O}(1)$ effect on the subimage EVPA:
\begin{align} \label{eq:mapproxEVPA}
    \EVPA_m \approx \EVPA_{m,{\rm Sch}}+(-1)^{m+1} \sin^{-1}(a/b_c)+\mathcal{O}\left(\frac{1}{r_{\rm s}}\right),
\end{align}
where $\EVPA_{m,{\rm Sch}}$ is the Schwarzschild face-on subimage ($m > 0$) EVPA. The critical impact parameter $b_c$ has a closed form expression in terms of $a,M$ (given by  e.g. Eq. (67) of  \cite{Gralla2020a}). Note that $\sin^{-1}(a/b_c)$ is approximated by $a/\sqrt{27}$ to within $8\%$ for all spins $-1\leq a\leq 1$, and $\sin^{-1}(a/b_c)|_{a=1}\approx 12^\circ$.

Like the approximation \eqref{eq:dchifaceon} for the direct image, \eqref{eq:mapproxEVPA} holds for any magnetic field configuration and boost parameter. The correction $\sin^{-1}(a/b_c)$ is analogous to the ``gravitational Faraday rotation" in the indirect image, thus extending the work of \cite{Balazs_1958,Pineault_1977,Fayos_1982,Ishihara_1988,Nouri_1999,Brodutch_2011} to the strongly lensed regime.

Unlike in the direct case, the effects of spin on the indirect image remain important even at large emission radii since the trajectory of a photon in the indirect image includes a radial turning point, which implies that the photon must move towards the black hole at some point after its emission into regions of higher curvature in which geometrical effects are important. 

\begin{figure*}[t]
    \centering
    \includegraphics[width=\textwidth]{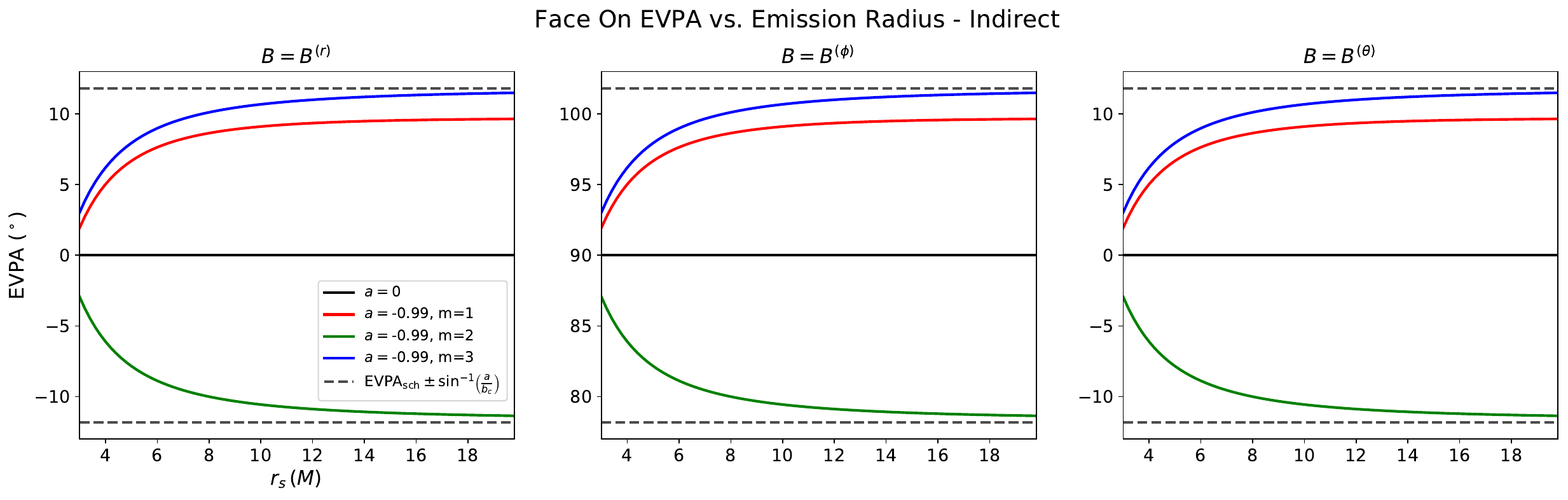}
    \caption{Face-on EVPA at $\varphi = 0$ as a function of emission radius (in units of $M$) for the first three sub-images. The columns are displayed for purely radial $(B = B^{(r)}, \eta_e = 0)$, toroidal $(B = B^{(\phi)},\eta_e = \pi)$, and vertical $(B = B^{(\theta)})$ magnetic fields. For  $a = -0.99$ the numerically computed $m =1,2,3$ subimage EVPAs are shown in red, green, and blue, respectively. The Schwarzschild EVPA is shown in black and the approximation $\EVPA_{\rm Sch}\pm \sin^{-1}(a/b_c)$ is shown with a dashed line.}
    
    \label{fig:faceoncompareindirect}
\end{figure*}

We demonstrate this effect in Figure \ref{fig:faceoncompareindirect}, which displays subimage EVPA as a function of emission radius (scaled in units of $M$) at $\varphi = 0$. Again, by axisymmetry, the EVPA at an arbitrary angle $\varphi = \vartheta$ is offset by exactly $\vartheta$ from the EVPA at $\varphi=0$. For maximal spin $a = -0.99$ the numerically computed $m =1,2,3$ subimage EVPAs are shown in red, green, and blue, respectively. The Schwarzschild EVPA, which does not depend on $m$ at fixed $\varphi$ for magnetic fields aligned with the local frame axes (see discussion of Figure~\ref{fig:faceoncomparedirect} in the previous subsection), is shown in black, and the approximation \eqref{eq:mapproxEVPA} of $\EVPA_{\rm Sch}\pm \sin^{-1}(a/b_c)$ is shown with a dashed black line. In Figure~\ref{fig:faceoncompareindirect}, the higher order subimages near a high-spin black hole oscillate in $m$ about the Schwarzschild EVPA before asymptoting to $\EVPA_{\rm Sch}\pm \sin^{-1}(a/b_c)$ at large $r_{\rm s}$ and $m$.\footnote{Note that at small radii, the EVPA of the $a=-0.99$, $m=1,2,3$ subimages shown in Figure~\ref{fig:faceoncompareindirect} is \textit{closer} to the Schwarzschild value than the corresponding $a=-0.99$ direct $m=0$ image shown in Figure~\ref{fig:faceoncomparedirect}. We have yet to find an intuitive explanation for this.} 

\section{Orbiting Hotspots} \label{sec:Hotspots}

An important application of our model is the computation of the polarization of an orbiting ``hotspot." As discussed in the introduction, such hotspots are frequently invoked to explain the observed flares in near-infrared and sub-millimeter frequencies near Sgr~A*. These flares can be tracked on short time-scales.

\subsection{Setup}

In our framework, the hotspot can be modelled as a point-source emitter orbiting on a circular, prograde, equatorial geodesic. Our simple model of point-source emission neglects any effects associated with the internal structure of the hotspot, such as shearing and cooling. Simulations of hotspots that include shearing and other plasma effects can be found in, e.g., \cite{Jeter_2020,Tiede_2020}. 

Measured from the ZAMO (which is not on a geodesic), the geodesic emitter has apparent boost (see \eqref{eq:betaboost} and e.g. Eq. (13) of \cite{Gates2020}): 
\begin{align} \label{eq:geodesicboost}
    \beta_v&= \frac{a^2-2 |a| \sqrt{r}+r^2}{\sqrt{a^2+r(r-2)} \left(|a|+r^{3/2}\right)}, \ \ \ \chi=-\pi/2,
\end{align}
where we explicitly include the absolute value to emphasize that under $a \to -a$, the boost remains in the same direction relative to the ZAMO. To track hotspot polarization, it is useful to work with the Stokes parameters $Q$ and $U$, defined following \cite{Himwich2020} by
\begin{align} \label{eq:StokesQU}
    Q&=(f^\beta)^2-(f^\alpha)^2,\qquad U=-2f^\alpha f^\beta.
\end{align}
As a reminder to the reader, $f^{\alpha}$ and $f^{\beta}$ are individual components of the screen polarization vector.
Note that our model of pure synchrotron radiation has a polarization fraction of $1$, i.e. $\sqrt{Q^2 + U^2} = I$. 

Tracking the motion of the hotspot in the $Q,U$ plane gives rise to polarization loops that trace the screen polarization pattern over time. The phase in $Q,U$ space tracks the EVPA via the relation 
\begin{equation}
{\rm EVPA}=\frac{1}{2}\arctan{\left(\frac{U}{Q}\right)},
\end{equation} 
following \eqref{eq:EVPA}, and the magnitude in $Q,U$ space tracks the polarized flux of the hotspot image. To convert intensity to flux, we multiply by a magnification factor that relates a differential area element on the observer's screen to a differential area element on the emitter's screen (described in detail in Appendix~\ref{app:magnification}).

Figures~\ref{fig:qupaneleq} and \ref{fig:qupanelvert} illustrate $Q,U$ loops produced by our semianalytic model for the intensity of direct emission of a hotspot in a purely vertical magnetic field and a purely equatorial field, respectively. Each panel of these figures shows spins of $a = 0, -0.99$ in red and blue, respectively, as well as the three inclinations of $\theta_o = 20^\circ, 45^\circ, 70^\circ$ in rows from top to bottom, and three emission radii of $r_{\rm s} = 6M, 8M, 10M$ in columns from left to right. Note that radii $r_{\rm s}\geq 6$ are outside (or at) the ISCO and admit stable orbits for all spins. The top left panel of each figure also indicates the location of four azimuthal emission coordinates $\phi$, spaced by $90^\circ$ on the Schwarzschild loop, computed using the fact that Schwarzschild geodesics lie in a plane and taking $\phi=\varphi=0$ aligned on the sky \cite{Narayan_2020}, \begin{align}
\label{eq:schphi}
    \tan(\varphi-n\pi)=\tan\phi\cos\theta_o,
\end{align}
where $n$ denotes subimage number (recall that $n=0$ has $m=1$ for $\beta>0$ and $m=0$ for $\beta<0$; see Footnote~\ref{foot:m}). 

\begin{figure*}[t]
    \centering
    \includegraphics[width=\textwidth]{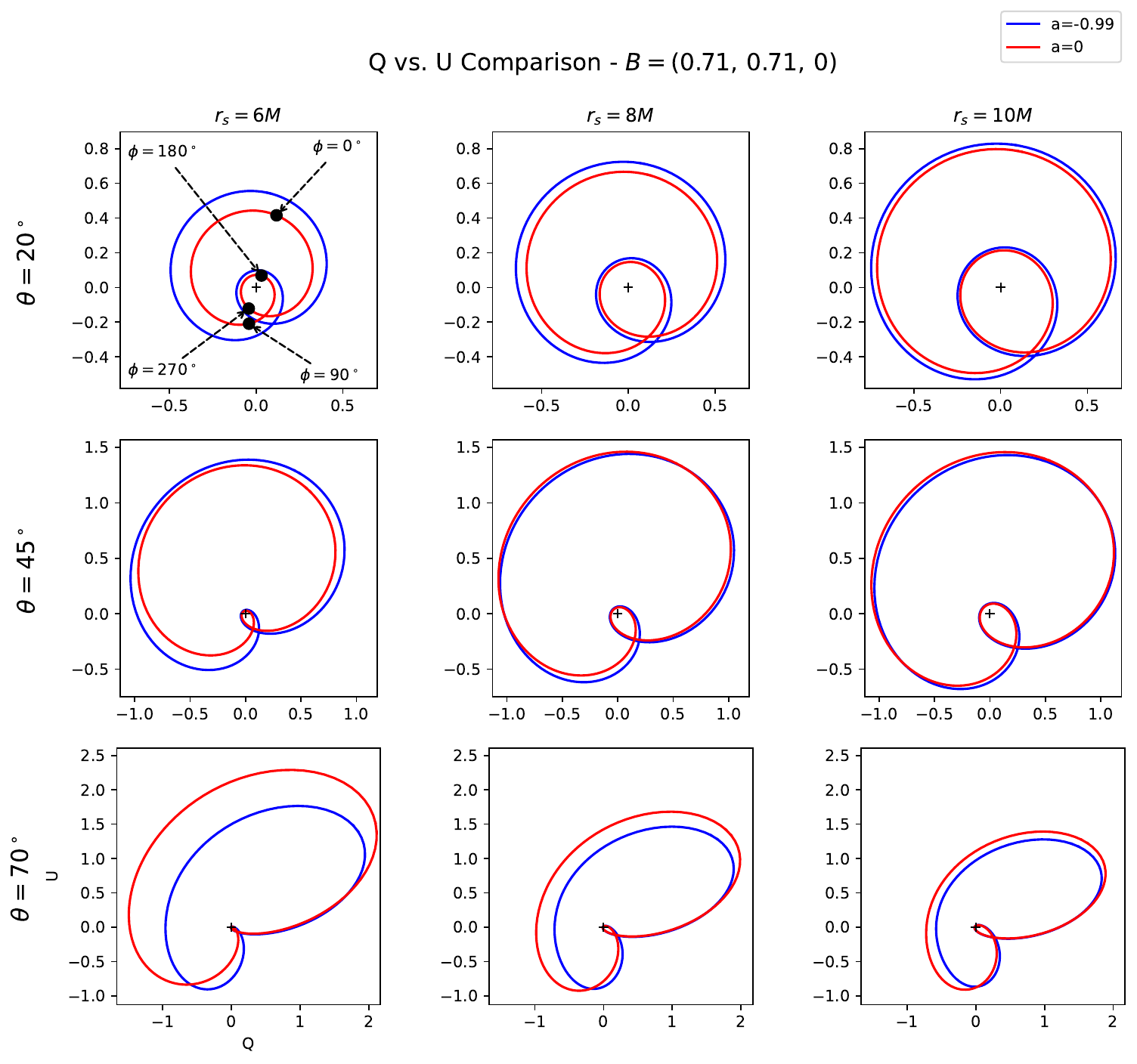}
    \caption{Polarized flux $Q$ vs. $U$ for equatorial magnetic fields. Plots are shown for spins of $a = 0, -0.99$ in red and blue, respectively, as well as three inclinations of $\theta_o = 20^\circ, 45^\circ, 70^\circ$ in rows from top to bottom, and three emission radii of $r_{\rm s} = 6M, 8M, 10M$ in columns from left to right.  The top left panel uses black dots to indicate four azimuthal emission coordinates $\phi$ on the Schwarzschild loop, spaced by $90^\circ$. Black crosshairs indicate the origin of each plot.}
    \label{fig:qupaneleq}
\end{figure*}
\begin{figure*}
    \centering
    \includegraphics[width=\textwidth]{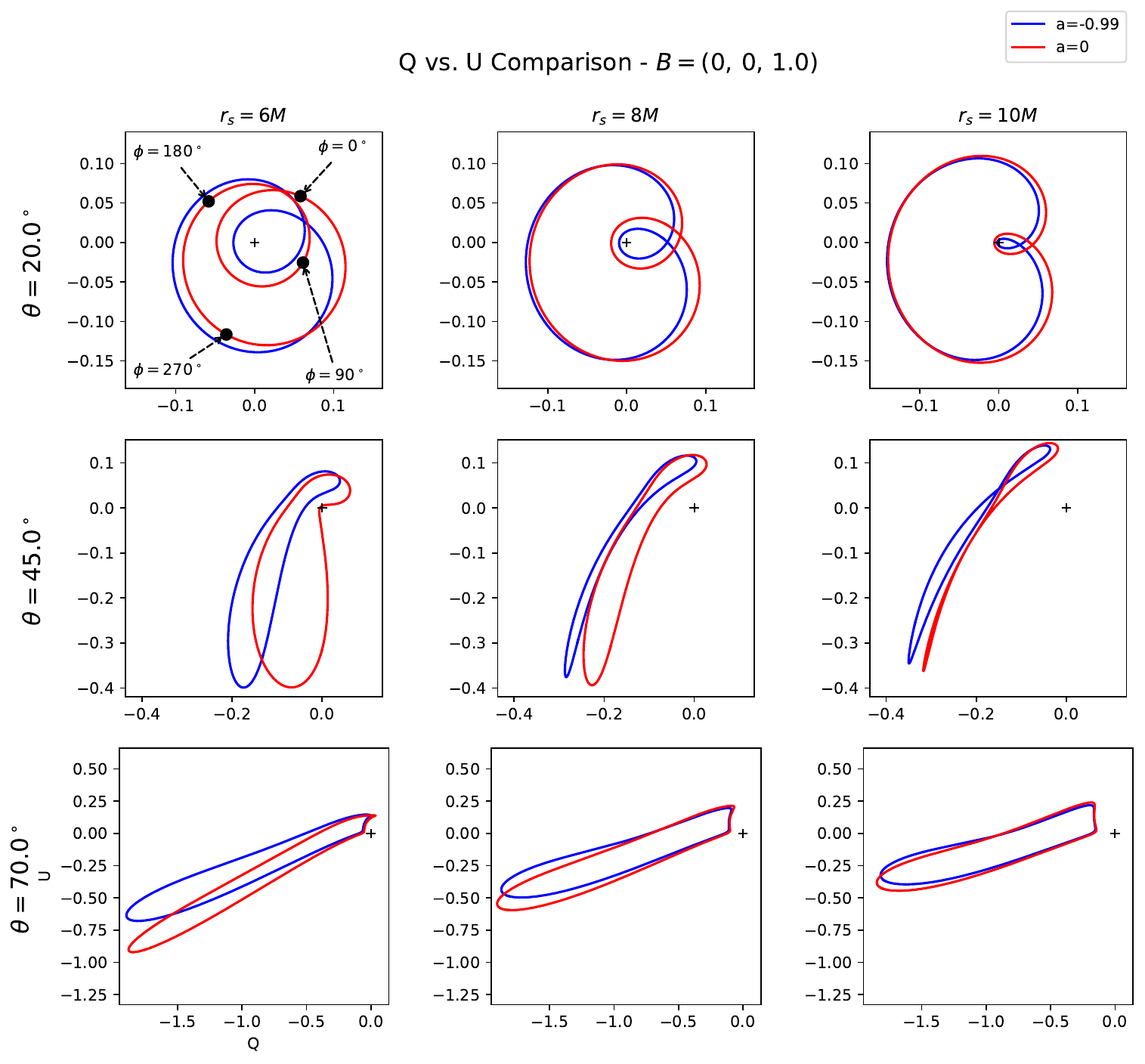}
    \caption{Polarized flux $Q$ vs. $U$ for vertical  magnetic field. Plots are shown for spins of $a = 0, -0.99$ in red and blue, respectively, as well as three inclinations of $\theta_o = 20^\circ, 45^\circ, 70^\circ$ in rows from top to bottom, and three emission radii of $r_{\rm s} = 6M, 8M, 10M$ in columns from left to right.  The top left panel uses black dots to indicate four azimuthal emission coordinates $\phi$ on the Schwarzschild loop, spaced by $90^\circ$. Black crosshairs indicate the origin of each plot.}
    \label{fig:qupanelvert}
\end{figure*}

As discussed in Sections~\ref{sec:Visualizations} and~\ref{sec:EVPA}, the differences between polarization emitted from an orbiter of fixed boost for high and low spin black holes is generally small at $r_{\rm s} \geq 6$, with the largest differences appearing at high inclination. For the geodesic orbiter \eqref{eq:geodesicboost} considered in this section, the boost parameter $\beta_v$  depends strongly on $a$, which is the primary source of differences seen in the $Q,U$ loops for low and high spins in Figures~\ref{fig:qupaneleq} and \ref{fig:qupanelvert}.

The $Q,U$ loops arising from equatorial and vertical magnetic field configurations have different topological structure. In particular, the equatorial field configuration gives rise to $Q,U$ plots with two loops enclosing the origin at all spins, emission radii, and inclination angles, while two loops are only present for low observer inclinations in the case of vertical fields. We explore the variations in $Q,U$ loop topology in more detail in the following section.

The $Q,U$ loop topology produced by our model, shown in Figures~\ref{fig:qupaneleq} and \ref{fig:qupanelvert}, is qualitatively consistent with the results of polarization loop studies for Sgr~A* by GRAVITY \cite{Gravity_2018,Gravity_2020}. In particular, we find that purely equatorial fields always give rise to two origin-enclosing $Q,U$ loops, while fields with a nonzero vertical component can give rise to a single loop.\footnote{Results of \cite{Gravity_2018,Gravity_2020} are displayed in intensity-normalized polarization fraction $Q/I,U/I$ and therefore cannot be quantitatively compared to our $Q,U$ (we assume a polarization fraction of $1$). However, normalization does not affect the observed phase in $Q,U$ space, so it is still meaningful to compare topological features of our loops.} Note that the double loops in the ray-traced and observed figures presented in \cite{Gravity_2018,Gravity_2020} do not encircle the origin. A shift in  observed or simulated $Q,U$ loops relative to the results of our simple geometric model of a point emitter could be due to the inclusion of indirect images, as we illustrate in Section \ref{subsec:QUsubimages}. A shift could also arise as a result of Faraday conversion of Stokes $Q,U$ to circularly polarized Stokes $V$ and could also arise for hotspots with physical extent large enough for $\vec{p} \times \vec{B}$ to vary significantly over the emitting region. Any stable polarized background will also add an offset (possibly $\phi$-dependent) in $Q,U$. 

The lack of strong dependence on spin in Figures~\ref{fig:qupaneleq} and \ref{fig:qupanelvert} parallels prior studies of both simulations and data. The work of \cite{Broderick:2006} ray-traced hot-spots around Sgr~A* in both NIR and radio frequencies and found minimal spin-dependence of the hotspot light curve for fixed orbital radius. Similarly, \cite{Gravity_2020} fit data for the July 22 Sgr~A* flare using three values of $a$ and saw minimal discrepancies.\footnote{While the polarization pattern of a point-like emitter orbiting on a fixed radius depends negligibly on the spin of the black hole, unpolarized observations of physical hotspots do provide an important observational target for the interferometric experiments that can constrain spin very well \cite{Tiede_2020}.}

\subsection{$Q,U$ Loop Topology} \label{subsec:QUtop}
\subsubsection*{Equatorial Fields}
For the equatorial fields $B^{(\theta)}=0$ shown in Figure~\ref{fig:qupaneleq}, the $Q,U$ loops in the direct emission all wrap twice around the origin, corresponding to $\psi_{\rm net}=2\pi$, where $\psi_{\rm net}$ is the net rotation of the EVPA over the course of one hotspot orbit. To explain this feature, we first note that for $\theta_o=0$, axisymmetry dictates that the $Q,U$ diagram will consist of two identical loops stacked on top of each other, both of which encircle the origin. Note that axisymmetry also implies that the EVPA will rotate in the same direction as the hotspot orbit. As $\theta_o$ increases, the two loops will become distinct. Suppose for contradiction that one loop were to disappear; by continuity, one of its points must pass through the origin $(Q=U=0)$. But this is impossible for $B^{(\theta)}=0$ since $p^{(z)}\neq 0$ implies that locally  $\sqrt{Q^2+U^2} \propto |\vec{B}\times \vec{p}|\neq 0$, which in the absence of Faraday conversion is directly proportional to the observed value.

Thus, two origin-enclosing $Q,U$ loops are always present in our geometric model with axisymmetric equatorial field configurations. In Figure~\ref{fig:qupaneleq}, the inner loop dramatically shrinks at high inclinations due to Doppler deboosting and small pitch angles (see \eqref{eq:sinzeta}), but always encircles the origin. With two nested loops present in the image, the direction of EVPA rotation matches the direction of motion of the hotspot on the sky. 

\subsubsection*{Vertical Fields}
In nonzero vertical fields $B^{(\theta)} \neq 0$, $Q,U$ loops can deform more dramatically. In particular, as the observer inclination $\theta_o$ increases, the inner loop may contract to a point and then vanish altogether, as demonstrated in each column of Figure~\ref{fig:qupanelvert} and also noted in \cite{Gravity_2020}. 

In images with two $Q,U$ loops, $\psi_{\rm net}=2\pi$, and the direction of EVPA rotation never changes over the hotspot orbit. In images with one $Q,U$ loop, the EVPA rotation direction instead changes twice over the hotspot orbit before the EVPA returns to its initial value.

At moderate inclinations, the $Q,U$ curve in Figure~\ref{fig:qupanelvert} at $\theta_o = 45^\circ$ with $r=10 M$ displays a lemniscate (figure-eight) shape that crosses over itself due to strong Doppler effects, but has the same EVPA rotation behavior as a single loop. In general, these lemniscates can form and then pull all the way through themselves as $\theta_o$ increases, producing the patterns seen in Figure~\ref{fig:qupanelvert} at $\theta_o = 70^{\circ}$.

\begin{figure}
    \centering
    \includegraphics[width=.49\textwidth]{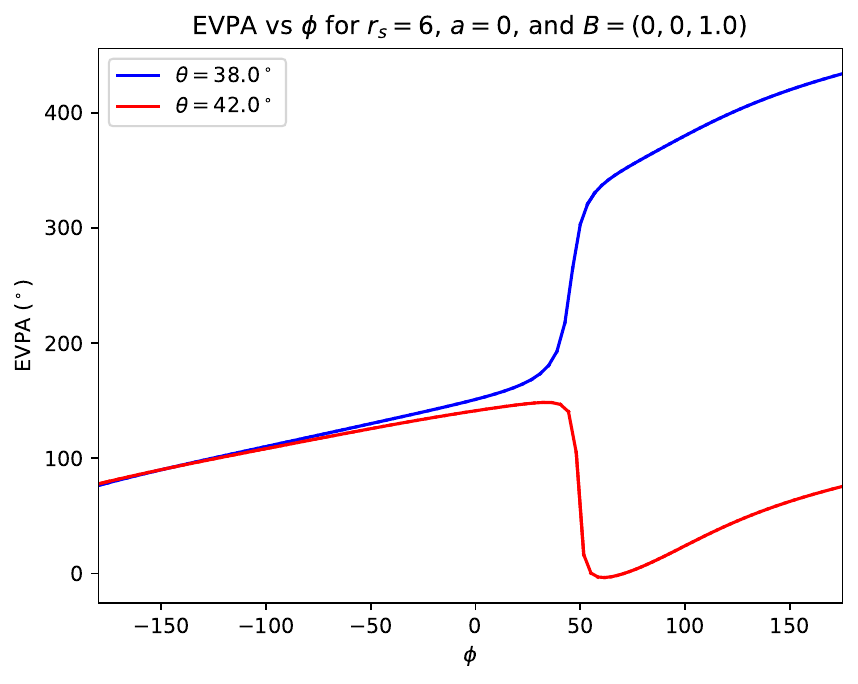}
    \caption{Schwarzschild EVPA as a function of azimuth $\phi$ for a hotspot orbiting in a purely vertical field. Two inclinations $\theta_o= 38^\circ, 42^\circ$ are shown in blue and red, respectively.}
    \label{fig:qucompare2}
\end{figure}

For illustration, in Figure~\ref{fig:qucompare2}, we plot the Schwarzschild EVPA as a function of orbital azimuth (again using \eqref{eq:schphi}) for a hotspot in a purely vertical field at two observer inclinations, $\theta_o=38^\circ, 42^\circ$. The EVPA continuously rotates in the same direction at $\theta_o=38^\circ$, corresponding to two $Q,U$ loops, while the EVPA changes rotation direction twice at $\theta_o=42^\circ$, corresponding to one loop. 

When considering \textit{only} the direct image in our model, the inner $Q,U$ loop always contracts to the origin as $\theta_o$ increases, as it must for $\psi_{\rm net}=2 \pi$ to change to $\psi_{\rm net}<180^\circ$; the single loop with a point at the origin corresponds to the boundary case of $\psi_{\rm net}=180^\circ$. We remark that the point to which the inner $Q,U$ loop contracts always appears as a cusp in $Q,U$ space, where the curvature of the $Q,U$ curve diverges. Though the cusp is smoothed out in physical observations with finite resolution, it is instructive to consider its geometric interpretation, included in Appendix~\ref{app:cuspangle}.

When the inner loop of a double-loop configuration is shifted off the origin, it no longer has $\psi_{\rm net}=2\pi$. But as long as the outer loop encircles the origin, then $\psi_{\rm net}\geq 180^\circ$. In \cite{Gravity_2018}, the observed single $Q,U$ loop for the July 28, 2018 flare had $\psi_{\rm net}\geq 180^\circ$, implying that it encircles the $Q,U$ origin. For the range of inferred radii, their simulations required a certain range of observer inclination $\theta_o \sim 15{-}30^{\circ}$ for a loop enclosing the origin to form (see \cite{Gravity_2018} Appendix D). This is consistent with our analysis: for $Q,U$ loops of direct images in vertical fields, we find loops enclosing the origin only at low-to-moderate inclinations. There is a \textit{single} origin-enclosing (and for us, origin-intersecting) loop for which the polarization and hotspot orbital periods are equal, only present at a (radius-dependent) critical inclination.  Physical effects, discussed in the previous and following subsections, that shift the loops would only function to modify these ranges.


We note additionally that for fields with zero azimuthal component (such as those that best fit the July 28 flare in \cite{Gravity_2020}), if $\psi_{\rm net}\leq 180^\circ$, then increasing $\theta_o$ will never cause $\psi_{\rm net}>180^\circ$. To prove this, note that for a fixed $a$, $r_{\rm s}$, and non-azimuthal $\vec{B}$ field, there is only one pair of conserved quantities $(\eta,\lambda)$ that satisfy $Q=U=0$ in the direct image, so there is only one inclination angle that satisfies $Q=U=0$ (see Appendix~\ref{app:cuspangle} for further justification). Hence loops of the direct image can pass through the origin at most once; if $\psi_{\rm net}\leq 180^\circ$, increasing $\theta_o$ cannot cause $\psi_{\rm net}>180^\circ$ again.

\subsection{Including Subimages} \label{subsec:QUsubimages}
Relative to the direct image, subimages appear both time-delayed and demagnified due to the longer path length of the corresponding photon trajectories. These effects are reviewed in Appendix \ref{app:timedelay} and \ref{app:magnification}, respectively, and must be taken into account to compute the contribution of indirect images to the polarized flux of an orbiting hotspot. 

\begin{figure}
    \centering
    \includegraphics[width=.49\textwidth]{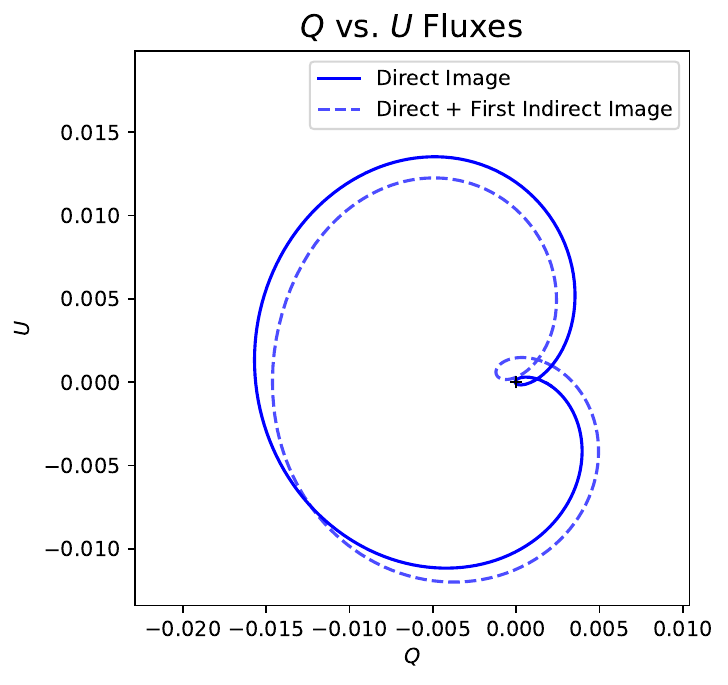}
    \caption{Polarized $Q,U$ flux loop for an orbiting hotspot at $r_{\rm s}=6,\,a=0,\,\theta_o=11^\circ,\,\vec{B}=\frac{1}{\sqrt{6}}(1,1,2)$, including the direct image (solid line) and the sum of the direct plus first indirect image (dashed line).}
    \label{fig:qusubimage}
\end{figure}

A sample polarized flux loop including the direct and first indirect image is displayed in Figure~\ref{fig:qusubimage} for a hotspot orbiting a Schwarzschild black hole with $r_{\rm s}=6, \theta_o=11^\circ$, and $\vec{B}=\frac{1}{\sqrt{6}}(1,1,2)$. As discussed in the previous subsection, the secondary loops in the direct emission are origin-enclosing. Figure~\ref{fig:qusubimage} illustrates that the time-delayed and demagnified $Q,U$ contribution from subimages can generally displace the secondary loop from encircling the origin. 

When the direct and indirect images are summed, this can cause depolarization; in this case, the depolarization will be strongest when the subimage has the largest relative flux \cite{Broderick:2006}. Future investigations may reveal whether depolarization in the total image could be a used as a robust signature of photon subrings, and could be an interesting future application of our model and traditional ray-tracing codes. 

\section{Summary} \label{sec:Summary} 

In this paper, we developed a semi-analytic toy model for polarized equatorial synchrotron  emission in the Kerr geometry. We computed the polarized images of axisymmetric fluid orbiting in various magnetic field configurations. Our simple model allowed us to isolate the individual effects of spin, magnetic field, and observer inclination on polarization signatures of the direct image and photon ring.

In the face-on case, we found that an analytic approximation to photon trajectories yields a  series expansion for the face-on EVPA, revealing a $\mathcal{O}(1/r_{\rm s}^2)$ correction from spin to the EVPA of the direct image but a $\mathcal{O}(1)$ correction to the EVPA of the indirect image. These calculations confirm results from previous studies of ``gravitational Faraday rotation'' \cite{Balazs_1958,Pineault_1977,Fayos_1982,Brodutch_2011,Ishihara_1988,Nouri_1999} and extend prior work to the strongly lensed regime. 

Our results bear direct relevance to observations and simulations of the supermassive black hole M87*, which is believed to be viewed at nearly face-on inclination  \citep{Walker_2018}. Our model suggests that differences between images of low-spin and high-spin simulations of M87* arise primarily from differences in accretion dynamics, as opposed to the frame-dragging of null geodesics. In the future, we hope to extend our calculations of analytic approximations to inclined observers as well. 

As an additional application of our toy model, we computed the polarized image of orbiting hotspots and provided an analytical framework for interpreting the topology of $Q,U$ loops observed in Sgr~A* flares. In particular, we explained how the magnetic field direction and observer inclination individually control the number and shape of $Q,U$ loops present, consistent with the interpretation provided by the GRAVITY collaboration \cite{Gravity_2018,Gravity_2020}.

While our toy models provide insight through simplicity, more complicated astrophysical effects are necessary to fully describe observed polarization patterns. For example, a realistic accretion model that includes turbulence cannot be modelled by axisymmetric rings of matter orbiting in a constant magnetic field, and Faraday effects can scramble the observed polarization pattern.
Our work does not attempt to provide a complete, physically realistic model; many highly-developed ray-tracing codes already exist for this purpose (such as  \cite{Dexter_2016,Moscibrodzka:2018,Younsi_2020}). Instead, our simple toy model provides intuition for the geometric effects of the Kerr spacetime on photon propagation and its imprint on polarized images. 

We have disentangled the roles of astrophysics and geometry, paving the way for a more complete understanding of polarimetric observations and how they are affected by a multitude of competing factors. Simple toy models have already played a key role in interpreting polarized black hole images (e.g., \cite{PaperVIII,Narayan_2020}), and we hope that our model and its future developments -- such as extensions to include non-equatorial emission and circular polarization -- will continue to provide new insights into polarimetric observations and simulations.

\section*{Acknowledgements}

We thank Avery Broderick, Dominic Chang, Andrew Chael, Delilah Gates, Ramesh Narayan, Angelo Ricarte, Paul Tiede, and Maciek Wielgus for useful discussions. We also thank Ziri Younsi for helpful comments on our manuscript. This work was supported in part by the Black Hole Initiative, which is funded by grants from the John Templeton Foundation and the Gordon and Betty Moore Foundation to Harvard University. We thank the National Science Foundation (AST-1716536, AST-1935980) and the Gordon and Betty Moore Foundation (GBMF-5278) for financial support of this work.

\appendix

\section{Details in Orbiting Fluid Model} \label{app:OrbiterDetails}

In this appendix, we include for concreteness the explicit expressions for the ZAMO tetrad and boost matrices described in Section \ref{sec:FluidModel}. The ZAMO tetrad appearing in \eqref{eq:tetradvector} can be arranged into a matrix 
\begin{equation}
    \left(e_{\rm {Z}}\right)^{\mu}_{(a)} = 
        \begin{pmatrix}
            \frac{1}{r_{\rm s}}\sqrt{\frac{\Xi_{\rm s} }{\Delta_{\rm s} }} & 0 & \frac{\omega_{\rm s}}{r_{\rm s}}\sqrt{\frac{\Xi_{\rm s}}{\Delta_{\rm s}}} & 0 \\
            0 & \frac{\sqrt{\Delta_{\rm s} }}{r_{\rm s}} & 0 & 0 \\
            0 & 0 & \frac{r_{\rm s}}{\sqrt{\Xi_{\rm s} }} & 0 \\
            0 & 0 & 0 & -\frac{1}{r_{\rm s}}, \\
        \end{pmatrix}
\end{equation}
so that
\begin{align}
    \left(
\begin{array}{c}
 V_{(t)} \\
 V_{(r)} \\
 V_{(\phi)} \\
 V_{(\theta)} \\
\end{array}
\right)&=e_Z\, \left(
\begin{array}{c}
 V_{t} \\
 V_{r} \\
 V_{\phi} \\
 V_{\theta} \\
\end{array}
\right).
\end{align}

The Lorentz transformation $\Lambda^{(a)}_{\ \ (b)}$ corresponding to \eqref{eq:betaboost} is given by the matrix
\begin{equation}
    \small
    \Lambda 
        = \begin{pmatrix}
             \gamma  & -\beta  \gamma  \cos \chi & -\beta  \gamma  \sin \chi & 0 \\
             -\beta  \gamma  \cos \chi & (\gamma -1) \cos ^2\chi+1 & (\gamma -1) \sin \chi \cos \chi & 0 \\
             -\beta  \gamma  \sin \chi & (\gamma -1) \sin \chi \cos \chi & (\gamma -1) \sin ^2\chi+1 & 0 \\
             0 & 0 & 0 & 1 \\
        \end{pmatrix}.
    \normalsize
\end{equation}

The explicit components of the tetrad \eqref{eq:boostedtetrad} relating the boosted local frame to Kerr are:
\begin{equation}
\begin{aligned}
    e'_{(t)}&=\frac{\gamma}{r_{\rm s}}\sqrt{\frac{\Xi_{\rm s}}{\Delta_{\rm s}}}\partial_t+\frac{\beta\gamma\cos\chi}{r_{\rm s}}\sqrt{\Delta_{\rm s}}\partial_r \\ &+\left( \frac{\gamma\omega_{\rm s}}{r_{\rm s}}\sqrt{\frac{\Xi_{\rm s}}{\Delta_{\rm s}}}+\frac{r_{\rm s}\beta\gamma\sin\chi}{\sqrt{\Xi_{\rm s}}}\right)\partial_\phi,\\
    e'_{(r)}&=\frac{\beta\gamma\cos\chi}{r_{\rm s}}\sqrt{\frac{\Xi_{\rm s}}{\Delta_{\rm s}}}\partial_t+\frac{1+(\gamma-1)\cos^2\chi}{r_{\rm s}}\sqrt{\Delta_{\rm s}}\partial_r \\
    &+ \left(\frac{\beta\gamma\omega_{\rm s}\cos\chi}{r_{\rm s}}\sqrt{\frac{\Xi_{\rm s}}{\Delta_{\rm s}}}+\frac{r_{\rm s}(\gamma-1)\cos\chi\sin\chi}{\sqrt{\Xi_{\rm s}}}\right)\partial_\phi,\\
    e'_{(\phi)} &= \frac{\beta\gamma\sin\chi}{r_{\rm s}}\sqrt{\frac{\Xi_{\rm s}}{\Delta_{\rm s}}}\partial_t +\frac{(\gamma-1)\cos\chi\sin\chi}{r_{\rm s}}\sqrt{\Delta_{\rm s}}\partial_r \\
    &+ \left(\frac{\beta\gamma\omega_{\rm s}\sin\chi}{r_{\rm s}}\sqrt{\frac{\Xi_{\rm s}}{\Delta_{\rm s}}}+\frac{r_{\rm s}((\gamma-1)\sin^2\chi+1)}{\sqrt{\Xi_{\rm s}}}\right)\partial_\phi,\\
    e'_{(\theta)}&=-\frac{1}{r_{\rm s}}\partial_\theta.
\end{aligned}
\end{equation}

\section{Details of Semi-Analytic Calculation} \label{app:AnalyticDetails}

\subsection{Definition of Special Functions} \label{app:Definitions}
The special functions in Section \ref{subsec:BardeenCoords} are defined: 
\begin{equation}
  \begin{aligned}
    F(x|k) &= \int_0^x \frac{d\theta}{\sqrt{1 - k\sin^2\theta}} = \int_0^{\sin{x}} \frac{dt}{\sqrt{(1 - t^2)(1 - kt^2)}}, \\
    K(k) &= F\left(\frac{\pi}{2}\big| k\right), \\
    \text{sn}(u|k) &= \sin\left( F^{-1}(x|k)\right),
  \end{aligned}
\end{equation} compatible with \textit{Mathematica} 12 \cite{Kapec2020}. Elliptic functions are computed in {\fontfamily{cmtt}\selectfont
python} using the package {\fontfamily{cmtt}\selectfont
mpmath} \cite{mpmath}. 

\subsection{Radial Roots} \label{app:RadialRoots}

See Appendix A of \cite{Gralla2020a}. We introduce
\begin{equation}
  \begin{aligned}
    \mathcal{A} &= a^2 - \eta - \lambda^2, \\
    \mathcal{B} &= 2M\left[\eta + (\lambda - a)^2 \right], \\
    \mathcal{C} &= - a^2\eta, \\
    \mathcal{P} &= - \frac{\mathcal{A}^2}{12} - \mathcal{C}, \\
    \mathcal{Q} &= - \frac{A}{3} \left[ \left(\frac{\mathcal{A}}{6}\right)^2 - \mathcal{C} \right] - \frac{\mathcal{B}^2}{8}, \\
  \end{aligned}
\end{equation}
and
\begin{equation}
  \begin{aligned}
    z &= \sqrt{\frac{-2( 3^{1/3}\mathcal{P})+2^{1/3}\mathcal{H}^{2/3}}{2( 6^{2/3}\mathcal{H}^{1/3})} - \frac{\mathcal{A}}{6}} > 0, \\
    \mathcal{H} &= -9\mathcal{Q}+\sqrt{12\mathcal{P}^3+81\mathcal{Q}^2}.
  \end{aligned}
\end{equation}
In terms of these definitions, the four roots are
\begin{equation}
  \begin{aligned}
    r_1 &= - z - \sqrt{-\frac{\mathcal{A}}{2} - z^2 + \frac{\mathcal{B}}{4z}}, \\
    r_2 &= - z + \sqrt{-\frac{\mathcal{A}}{2} - z^2 + \frac{\mathcal{B}}{4z}} ,\\
    r_3 &= z - \sqrt{-\frac{\mathcal{A}}{2} - z^2 - \frac{\mathcal{B}}{4z}}, \\
    r_4 &= z + \sqrt{-\frac{\mathcal{A}}{2} - z^2 - \frac{\mathcal{B}}{4z}}. \\
  \end{aligned}
\end{equation}
Note that we have altered the root of the resolvent cubic from \cite{Gralla2020b} so that all cube roots can be taken as principal (i.e. we take $(-1)^{1/3}=e^{i\pi/3}$ as opposed to $-1$). This convention is default in \textit{Mathematica} 12 and {\fontfamily{cmtt}\selectfont
numpy}.

\subsection{Sign $\pm_r$ of Initial Radial Motion} \label{app:Rmotion}

The sign $\pm_r$ in \eqref{eq:fourmom} depends on $\{\lambda,\eta, r_{\rm s}, a, \theta_o,m\}$ and must be computed semi-analytically. To do so, one must first check whether a given geodesic contains a radial turning point (equivalently, whether the geodesic terminates inside or outside the critical curve \cite{Gralla2020a}). Geodesics that do not contain a turning point (and thus fall inside the critical curve) must have $\pm_r=+1$ to reach infinity. For geodesics that do contain a turning point, one must check whether a ray shot backwards encounters the turning point before or after the desired emission coordinates. For geodesics with a turning point, the radial integral $I_r$ has antiderivative given by (A9) of \cite{Gralla2020b}:
\begin{align} \label{eq:antideriv}
    \int \frac{dr}{\sqrt{\mathcal{R}(r)}}&= \frac{2}{\sqrt{r_{31}r_{42}}}F\left(\arcsin{\sqrt{\frac{r-r_4}{r-r_3}\frac{r_{31}}{r_{41}}}}\bigg|k\right).
\end{align}
The turning point is located at $r_4$, at which point the antiderivative vanishes. Following \cite{Gralla2020a}, the radial integral $I_{r}^{\rm turn}$ for the portion of the geodesic connecting the turning point to the observer is simply the limit of Eq. \eqref{eq:antideriv} with $r\to\infty$, as we assume the observer is located at infinity,\begin{align} \label{eq:Irturn}
    I_{r}^{\rm turn}&=\frac{2}{\sqrt{r_{31}r_{42}}}F\left(\arcsin{\sqrt{\frac{r_{31}}{r_{41}}}}\bigg|k\right).
\end{align}

Since $I_r=G_\theta$ is strictly increasing along the geodesic \cite{Gralla2020a}, then rays with $G_\theta^m<I_{r}^{\rm turn}$ will not encounter the turning point, and rays with $G_\theta^m>I_{r}^{\rm turn}$ will encounter the turning point. We therefore have\begin{align}
    \pm_r &= \begin{cases} 
      +1 & b< b_c \\
      \sign(I_{r}^{\rm turn}-G^m_\theta) & b>b_c.
   \end{cases}
\end{align}
As $m$ grows, the boundary between rays with $\pm_r=1$ and $\pm_r=-1$ approaches the critical curve; rays with $m\gg 1$ that terminate outside the critical curve must be emitted with $\pm_r=-1$ so that they can asymptote to a spherical photon orbit before escaping to infinity. Thus $\pm_r\to \sign(b-b_c)$ as $m\to\infty$. 

\subsection{Critical Parameters and Critical Curve} \label{app:criticalcurve}

We follow \cite{Gralla2020a}, to which we refer readers for more details and discussion. Tildes denote ``critical" parameters, evaluated at radii $\tilde{r}$ at which the Kerr geometry admits bound photon orbits, lying within  $\tilde{r}_- \leq \tilde{r} \leq \tilde{r}_+$, with
\begin{equation}
\tilde{r}_{\pm} = 2 M \left[ 1 + \cos \left(\frac{2}{3} \arccos\left(\pm \frac{a}{M}\right)\right) \right].
\end{equation}
Photons on orbits with radius $\tilde{r}$ have critical parameters 
\begin{equation}
\begin{aligned}
\tilde{\lambda} &= a + \frac{\tilde{r}}{a} \left[ \tilde{r} - \frac{2\tilde{\Delta}}{\tilde{r} - M}\right], \\
\tilde{\eta} &= \frac{\tilde{r}^3}{a^2}\left[ \frac{4M\tilde{\Delta}}{(\tilde{r} - M)^2} - \tilde{r} \right].
\end{aligned}
\end{equation}
On the observer screen, these parameters define a closed ``critical curve" parametrized by $\tilde{r}$. Note that each radius $\tilde{r}$ defines two points on the critical curve corresponding to the two possible values $\pm_o$ (two possible values $\pm \beta$). 

\subsection{Schwarzschild Subimage Time Delay }
\label{app:timedelay}
Due to longer path lengths between the source and observer, subimages of an emitting source appear at a time delay relative to the direct image. The expression for time elapsed between source and observer follows from the geodesic equation (see e.g. (7c) of \cite{Gralla2020a}), and in the Schwarzschild $(a\to0)$ case is given by 
\begin{align}
\label{eq:timeint}
    \Delta t_{n} = t_{o,n}-t_{s,n}&=\fint dr\,\frac{r^3}{(r-2)\sqrt{\mathcal{R}(r)}},
\end{align}
where $t_{s,m}$ and $t_{o,m}$ denote the times of source emission and observation, respectively, for the $n$-th subimage. To incorporate time delay between the direct and indirect image in the hotspot model, we suppose that the hotspot is located at $\phi=0$ when $t_{s}=0$. In the $a=0$ case, we may use \eqref{eq:schphi} to obtain
\begin{align} \label{eq:timedelay}
     \tan(\varphi-n\pi)=\tan[\omega_{\rm s}\, (t_{o,n}-\Delta t_{n})]\cos\theta_o,
\end{align}
where $\omega_{\rm s} =-r_{\rm s}^{-3/2}$ is the angular velocity of the equatorial geodesic in Schwarzschild \cite{Gralla2018} (corresponding to \eqref{eq:geodesicboost} with $a=0$), and we have added the minus sign to be consistent with clockwise motion on the sky ($\chi=-\pi/2$). Using (\ref{eq:timeint}-\ref{eq:timedelay}), we numerically compute $t_{o,n}(\varphi)$, and interpolate the inverse function $\varphi(t_{o,n})$ to find $Q(\varphi(t_{o,n})), U(\varphi(t_{o,n}))$ as a function of $t_{o,n}$.  We sum the Stokes parameters of distinct subimages arriving at the same $t_o$ to obtain the total observed intensity or flux.

\subsection{Magnification and Flux}
\label{app:magnification}
 
To compute the image flux, we must consider a bundle of null geodesics with infinitesimal but nonzero area emanating from the emitting source. The total flux from such a bundle of null rays is the integral of the observer intensity over the projection of the area of the bundle of rays on the observer's screen. In practice, this involves computing the Jacobian relating differential area elements between the emitter and observer screen, which is outlined in e.g. \cite{Cunningham1973,Gralla2018}. 
The area element of the solid angle corresponding to the bundle of rays of the observer screen at radius $r_o$ in Bardeen coordinates is $d \alpha d \beta /r_o^2$, so the Stokes linearly polarized flux is the area integral
\begin{equation} \label{eq:flux}
\oiint \frac{d \alpha \ d \beta}{r_o^2} \left(Q,U\right) = \frac{\left(Q,U\right)}{r_o^2}\mathcal{A}, \ \ \ \ \mathcal{A} = \oiint d \alpha \ d \beta ,
\end{equation}
where $\mathcal{A}$ is the area of the image on the observer screen, and $Q,U$ are defined using \eqref{eq:PolObserved} and \eqref{eq:StokesQU}, including the redshift but \textit{without} including the path length, as the hotspot is an isotropic circular emitter rather than a thin disk. In \eqref{eq:flux} we take $Q,U$ (along with $r_o$) to be constant over the image of an infinitesimal bundle of rays. 
We compute the area $\mathcal{A}$ directly following \cite{Gralla2018}, defining the local ``source screen" $(t) = (r) = 0$ in the frame of the orbiting geodesic emitter (corresponding to boost \eqref{eq:geodesicboost}). Note that the factor of $r_o^2$ drops out of the ratio of direct and indirect image flux.

\section{Image Symmetries}
\label{app:symmetries}
This appendix records changes in the polarized image under various parameter sign flips. Such transformations correspond to symmetries of geodesics, which only in certain fluid and magnetic field configurations correspond to symmetries of the polarized image. 

\begin{figure*}
    \centering
    \includegraphics[width=\textwidth]{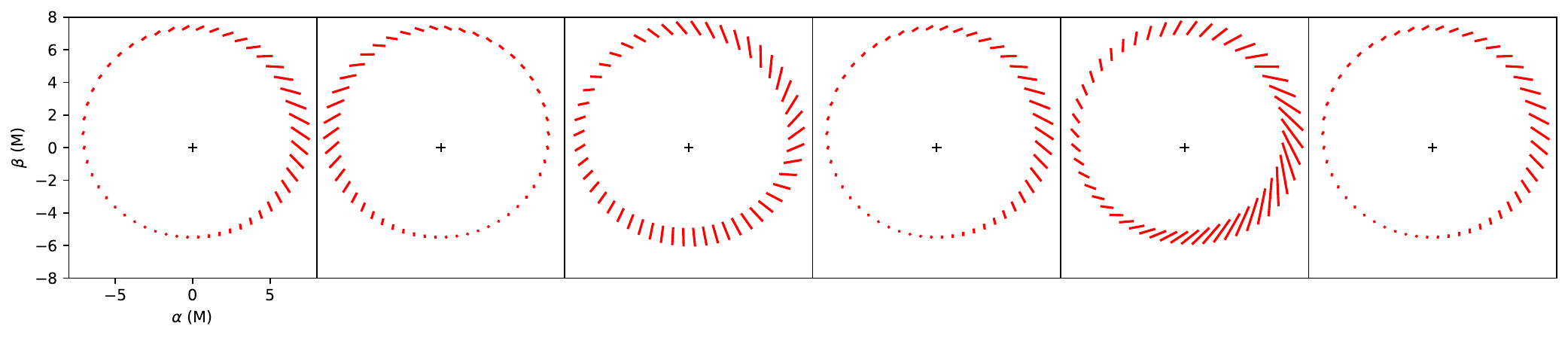}
    \caption{Sample images produced under various transformations of the ``original image" in the leftmost panel, aligned with the corresponding symmetry transformation in Table \ref{tab:symtable}. The original image has  $r_{\rm s}=6,\,a=-0.5,\,\theta_o=30^\circ,\,\vec{B}=\frac{1}{\sqrt{3}}(1,1,1)$, and boost parameters corresponding to a prograde geodesic (\ref{eq:geodesicboost}).}
    \label{fig:symmetryfig}
\end{figure*}

\begin{table*} 
\centering
\hspace*{.5cm}\begin{tabular}{|C{2.74cm}|C{2.74cm}|C{2.74cm}|C{2.74cm}|C{2.74cm}|C{2.74cm}|}
\hline
  & \begin{tabular}{@{}c@{}} $a,\alpha$ \\ $\downarrow$ \\ $-a,-\alpha$ \end{tabular}& \begin{tabular}{@{}c@{}} $a,\alpha,\chi$ \\ $\downarrow$ \\ $-a,-\alpha,-\chi$ \end{tabular} &  \begin{tabular}{@{}c@{}} $a,\alpha,\chi, B^{(\phi)}$ \\ $\downarrow$ \\ $-a,-\alpha,-\chi, -B^{(\phi)}$ \end{tabular} & \begin{tabular}{@{}c@{}} $\theta_o,\beta$ \\ $\downarrow$ \\$\pi-\theta_o,-\beta$\end{tabular} & \begin{tabular}{@{}c@{}} $\theta_o,\beta,B^{(\theta)}$  \\ $\downarrow$ \\ $\pi-\theta_o,-\beta,-B^{(\theta)}$ \end{tabular}\\ \hline
 Arrival Position & \checks & \checks & \checks & \checks & \checks\\\hline
  Doppler Boost & \x & \checks & \checks & \checks & \checks\\\hline
 $I,Q,U,V$ & \x & \x & \checks & \x & \checks \\\hline
\end{tabular}
\caption{Effects of various parameter transformations on geodesic arrival position, Doppler boost factor, and the Stokes parameters $I,Q,U,V$ in our model of equatorial emission. A check \checks $\,$ ($\,$\x   $\,$) indicates that a quantity is (is not, generically) invariant under the given transformation. }\label{tab:symtable}
\end{table*}

\subsubsection*{Observer Inclination $\theta_o \to \pi - \theta_o$}

Consider a photon arriving at position $(\alpha, \beta)$ for an observer at inclination $\theta_o < \frac{\pi}{2}$ relative to the positive black hole spin axis. The observer at location $\pi - \theta_o$ will observe photons with an opposite sign momentum along the spin axis, $\pm_o \to \mp_o$, and the arrival position in $\beta$ will correspondingly flip: $\beta \to - \beta$. By inspection of \eqref{eq:Bardeeninverse} the geodesic conserved quantities are invariant.\footnote{The direct image observed at inclination $\pi - \theta_o$ will thus have $m=0$ for photons arriving on the top half of the image and $m=1$ for photons arriving on the bottom half of the image. Note that Eq. 82 of \cite{Gralla2020a} generalizes to $\overline{m}=m-H(\beta\cos\theta_o)$, meaning the direct image with $\overline{m}=0$ is unchanged by the transformation $\theta_o\to\pi-\theta_o,\beta\to-\beta$.}

For the polarized image with a non-vertical field ($B^{(\theta)}=0$), the sign flip $\pm_o$ will cause $f^{(r)}$ and $f^{(\phi)}$ to flip sign by inspection of \eqref{eq:crossprod}. In this case, $\kappa_1$ flips sign and the polarization corresponding to $(f^{\alpha},f^{\beta})$ at $(\alpha, \beta)$ for an observer at $\theta_o < \frac{\pi}{2}$ will appear as $(-f^{\alpha},f^{\beta})$ at position $(\alpha, -\beta)$ for an observer at $\pi - \theta_o$. 
For fields that contain nonzero $B^{(\theta)}$, the same result is produced under the additional transformation $B^{(\theta)}\to -B^{(\theta)}$.

Furthermore, an observer at $\pi - \theta_o$ with flipped values of the equatorial magnetic field $\vec{B}_{\rm eq}\to -\vec{B}_{\rm eq}$ corresponds to a flip $f^\theta\to-f^\theta$, which complex conjugates $\kappa \to \bar{\kappa}$. With this flip in fields, the polarization corresponding to $(f^{\alpha},f^{\beta})$ at $(\alpha, \beta)$ for an observer at $\theta_o < \frac{\pi}{2}$ will appear as $(f^{\alpha},-f^{\beta})$ at position $(\alpha, -\beta)$ for an observer at $\pi - \theta_o$. 

Since the polarization on the screen is defined up to an overall sign, in all of the cases above we find that the polarized image will be reflected across the $\alpha$ axis. Note also that at $\pi - \theta_o$, the \textit{apparent} direction of clockwise/counterclockwise fluid motion flips. 

Finally, we note that $\theta_o \to - \theta_o$ corresponds to a rotation in $\phi_o$, which will not change the image position of axisymmetric configurations or the polarization pattern of axisymmetric magnetic field configurations. 

\subsubsection*{Spin Direction $a \to -a$}

Consider again a photon arriving at position $(\alpha, \beta)$ for an observer at inclination $\theta_o < \frac{\pi}{2}$ relative to the positive black hole spin axis, which is defined as $\hat{z}$. If the spin flips sign, its angular momentum changes to be aligned with $-\hat{z}$, so the definition of $\hat{\beta}$ changes orientation and the observer will be at an angle $\pi - \theta_o$ relative to the spin vector. In this new coordinate system, rotation in $\phi$ changes from counterclockwise to clockwise on the screen, and the image that appeared at $\alpha \hat{\alpha} + \beta\hat{\beta}$ in the old coordinates will appear at $(- \alpha)\hat{\alpha} + (-\beta) (- \hat{\beta})$ in the new coordinates , i.e., with $\alpha$ flipped relative to the old image position. By inspection of \eqref{eq:Bardeeninverse}, $\lambda$ flips sign and $\eta$ is invariant; thus $p^{\phi}$ flips sign. 

For the polarized image of \emph{unboosted} ZAMO emission, if $B^{(\phi)} = 0$, the sign flip in $p^{\phi}$ will cause $f^{(r)}$ and $f^{(\theta)}$ to flip sign by inspection of \eqref{eq:crossprod}. In this case, $\kappa_1$ flips sign and $\mu$ also flips sign, so the polarization $(f^\alpha,-f^\beta)$ at $(\alpha, \beta)$ for an observer of spin $+a$ aligned with $+\hat{z}$ will appear as the polarization $(f^\alpha,f^\beta)$ to an observer with spin aligned $-a$ at position $(-\alpha, \beta)$ in the original coordinates. For fields with $B^{(\phi)}\neq 0$, the same result is produced under the additional transformation $B^{(\phi)}\to -B^{(\phi)}$.

Furthermore, the transformation with unboosted emission  $a,B^{(r)},B^{(\theta)}\to-a,-B^{(r)},-B^{(\theta)}$ corresponds to a flip $f^\phi\to-f^\phi$, which complex conjugates $\kappa\to\overline{\kappa}$. Hence, under this transformation, the polarization corresponding to $(f^\alpha,f^\beta)$ at screen coordinates $(\alpha,\beta)$ and spin $+a$ aligned with $+\hat{z}$ will appear as polarization $(-f^\alpha,f^\beta)$ to an observer with spin aligned $-a$ at position $(-\alpha,\beta)$.

Since the EVPA is defined up to a sign in $\pm(f^{\alpha},f^{\beta})$, for all the $a \to -a$ cases above, the polarized image will be reflected across the $\beta$ axis. Note also that with $a \to -a$, the direction of ZAMO motion changes direction and will appear clockwise on the observer screen, with the boost $\chi$ now in the opposite direction of the ZAMO motion. The transformation $a,\chi\to -a,-\chi$ leaves the fluid frame tetrad $e_{(a)}^{\,\mu}$ invariant, so the above transformations for $a\to-a$ apply to boosted emission as well if the azimuthal boost direction flips $\chi \to -\chi$.

\subsubsection*{Combined Symmetries}
Since the polarization symmetries for $\theta_o\to\pi-\theta_o$ reflect the image across the $\alpha$ axis while the polarization symmetries for $a,\chi\to-a,-\chi$ flip the image across the $\beta$ axis, the two sets of transformations will produce images that are rotated by $\pi$ relative to each other. 

We note additionally that $Q$ and $U$ are invariant under a symmetry transformation if and only if the arrival positions of the geodesics are unchanged and the pitch angle~\eqref{eq:sinzeta} is unchanged, which is equivalent to preserving $I$. Hence $I,Q,U$ all obey the same set of symmetry relations.

The effect of a variety of the symmetry transformations on arrival position, boost factor, and polarization presented in this section are listed in Table~\ref{tab:symtable}. A sample image with $r_{\rm s}=6,\,a=-0.5,\,\theta_o=30^\circ,\,\vec{B}=\frac{1}{\sqrt{3}}(1,1,1)$, and boost parameters corresponding to a prograde geodesic (\ref{eq:geodesicboost}) is explicitly transformed to showcase these symmetries in Figure~\ref{fig:symmetryfig}.

\subsubsection*{Circular Polarization}

Our model can naturally be extended to include circular polarization, which is encoded in Stokes $V$ and is invariant on  geodesics in the absence of Faraday effects. At the source, $V$ is a complicated function of $\vec{B}$ such that the sign of the magnetic field components enters only into the sign of $V$ via the relation ${\rm sign}(V) = {\rm sign}(\vec{p}\cdot\vec{B})$ \cite{Ricarte:2021frd}. With this relation, we can derive the effect of the same symmetry transformations, which only affect $V$ through sign$(V)$ (assuming the circular polarization is entirely intrinsic). As previously stated, the transformation $a,\alpha,\chi\to -a,-\alpha,-\chi$ flips the sign of $p^{(\phi)}$, so taking $a,\alpha,\chi,B^{(\phi)}\to -a,-\alpha,-\chi,-B^{(\phi)}$ leaves $\vec{p}\cdot \vec{B}$ and hence sign$(V)$ and $V$ unchanged. Similarly, the transformation $\theta_o,\beta\to\pi-\theta_o,-\beta$ flips the sign of $p^{(\theta)}$, so taking $\theta_o,\beta, B^{(\theta)}\to\pi-\theta_o,-\beta,-B^{(\theta)}$ also leaves sign$(V)$ and $V$  unchanged. 

These symmetries are also listed in Table~\ref{tab:symtable}. The other transformations listed in Table~\ref{tab:symtable} flip the sign of only certain components of the dot product $\vec{p}\cdot\vec{B}$, so the effect on sign$(V)$ cannot be generically determined.

\subsubsection*{Universal Subimage Symmetries}

Our result~\eqref{eq:mapproxEVPA} is also consistent with the subimage polarization symmetry relations derived in \cite{Himwich2020}. In particular, suppose that $\vec{B}=B^{(\theta)}$ so that equatorial sources have $f_m^\theta=-f_{m+1}^\theta$. Together with $p_m^\theta=-p_{m+1}^\theta$, this implies that $\kappa$ is complex conjugated across subrings (as shown in \cite{Himwich2020}). Then Eqs.~\eqref{eq:kappadef}, \eqref{eq:ffaceon}, and \eqref{eq:mapproxEVPA} imply\begin{align}
\nonumber\frac{f^\alpha_{m}{+}f^\alpha_{m+1}}{f^\beta_{m}{+}f^\beta_{m+1}}&\propto \frac{\cos{({\rm EVPA}_{\rm sch})} \frac{a}{b_c}}{\sin{({\rm EVPA}_{\rm sch})} \frac{a}{b_c}}=\cot({\rm EVPA}_{\rm sch}),
    \\ 
   \nonumber \frac{f^\beta_{m}{-}f^\beta_{m+1}}{f^\alpha_{m}{-}f^\alpha_{m+1}}&\propto\frac{\pm(-1)^{m}\cos{({\rm EVPA}_{\rm sch})} \sqrt{1{-}\frac{a^2}{b_c^2}}}{\mp(-1)^m\sin{({\rm EVPA}_{\rm sch})} \sqrt{1{-}\frac{a^2}{b_c^2}}} , \\
   &\qquad =-\cot({\rm EVPA}_{\rm sch}),\\
\end{align}
Noting that 
\begin{equation}
\cot({\rm EVPA}_{\rm sch}) = \cot\varphi,
\end{equation}
the result agrees with Eqs.~(36) and (37) of \cite{Himwich2020}. In re-deriving these relations, we have used the additional fact that the screen polarization vector $(f^{\alpha},f^{\beta})$ flips sign across subrings for the specific case of $a=0$, $B=B^{(\theta)}$.

\section{$Q,U$ Cusp Formation in Vertical Fields} \label{app:cuspangle}

\begin{figure*}[t]
    \centering
    \includegraphics[width=\textwidth]{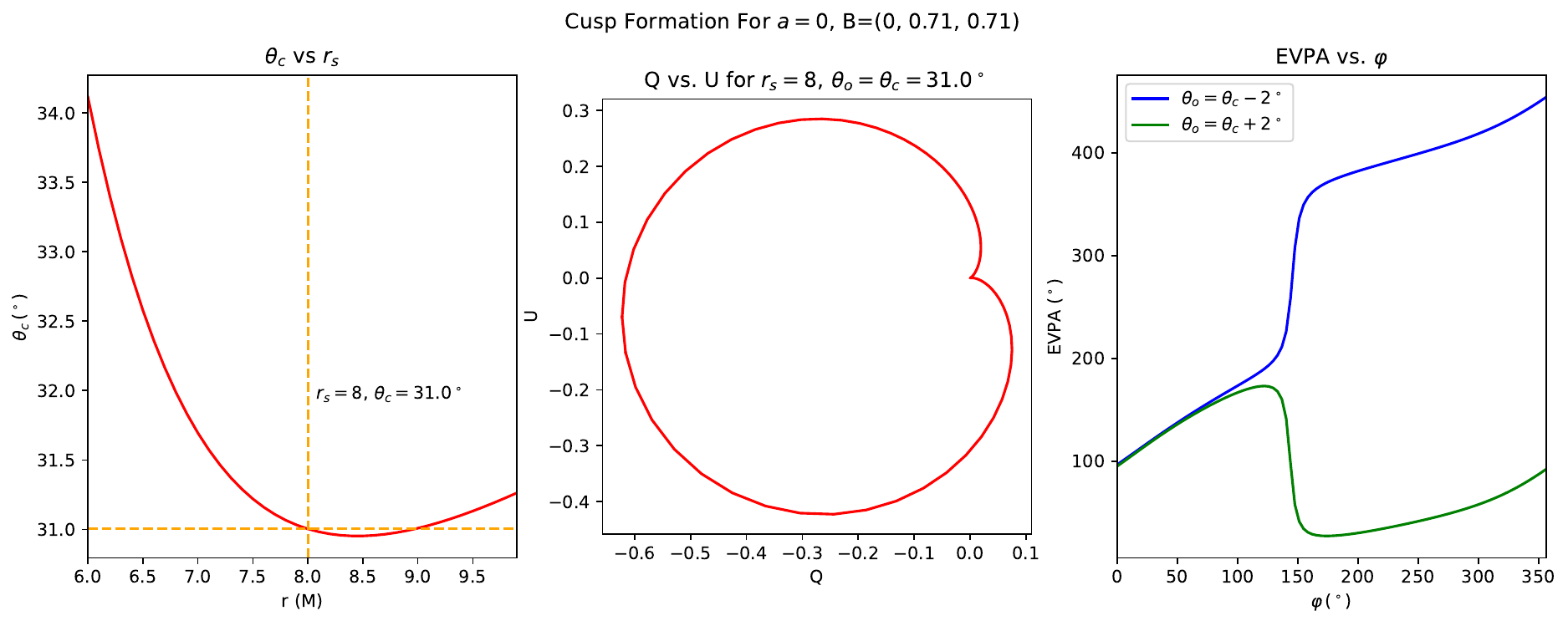}
    \caption{Polarization at $\theta_{\rm cusp}$ for a hotspot orbiting in a magnetic field $\vec{B}=\frac{1}{\sqrt{2}}(0,1.0,1.0)$ around a Schwarzschild black hole. Left panel: $\theta_{\rm cusp}$ as a function of emission radius $r_{\rm s}$, with dashed lines intersecting at $r_{\rm s} = 8$, $\theta_{\rm cusp} = 32.95^{\circ}$. Center panel: polarization $Q,U$ loop at $r_{\rm s} = 8$, $\theta_{\rm cusp} = 32.95^{\circ}$. Right panel: EVPA as a function of orbital azimuth for observer inclinations of $\theta_o=\theta_{\rm cusp} \pm 2^\circ$ in green and blue, respectively.}
    \label{fig:cuspplot}
\end{figure*}

This appendix discusses an interesting feature of the $Q,U$ loop topology (Section~\ref{subsec:QUtop}) for vertical fields displayed in Figure~\ref{fig:qupanelvert}: as $\theta_o$ increases, the point to which the inner loop of the direct image contracts always appears as a cusp in $Q,U$ space, where the curvature of the $Q,U$ curve diverges. The cusp will not appear in physical observations with finite resolution, but it is still instructive to examine the geometric origin of $Q,U$ loop cusp formation. 

To gain intuition for this phenomenon, note as in Section~\ref{subsec:QUtop} that for two $Q,U$ loops, one has $\psi_{\rm net}=2\pi$ with constant rotational direction, while for one $Q,U$ loop, $\psi_{\rm net}\neq 2\pi$; in the latter case, the EVPA rotation direction changes twice over the orbit before the EVPA returns to its initial value. As the observer inclination increases, a cusp forms at $\theta_{\rm cusp}$ at the boundary between these two types of EVPA behavior. 

Figure~\ref{fig:qucompare2} plots the Schwarzschild EVPA as a function of orbital azimuth for a hotspot in a purely vertical field at two observer inclinations $\theta_o=42^\circ, 38^\circ$ slightly above and below $\theta_{\rm cusp}$. As notes in Section~\ref{subsec:QUtop}, the EVPA continuously rotates in the same direction at $\theta_o=38^\circ$, corresponding to two $Q,U$ loops, while the EVPA changes rotation direction twice at $\theta_o=42^\circ$, corresponding to one $Q,U$ loop. At $\theta_{\rm cusp}$, which lies between these two values, the EVPA changes discontinuously, i.e. its derivative becomes singular, as Figure~\ref{fig:qucompare2} suggests. In the case of Figure~\ref{fig:qucompare2}, $\theta_{\rm cusp} = 40.7^\circ$, which can be calculated explicitly as described below.

As noted above, when an origin-enclosing loop contracts, by continuity one of its points must pass through the origin $(Q=U=0)$. For vertical fields, we observe that the cusp formed when a loop contracts is always located at the origin, where the EVPA is not well-defined (i.e. rotates infinitely quickly). The $Q,U$ cusp and the $Q,U$ origin both reflect a singularity in EVPA rotation, and we intuitively expect that they coincide, i.e. that there is always a cusp in the $Q,U$ loop if $Q = U = 0$.\footnote{We are able to prove this explicitly in the specific case of a purely vertical field. We expect that we should be able to extend the proof to any field configuration with nonzero field, and that the converse also holds. We leave this to future work.} 

Thus, given $\vec{B}$, $a$, and $r_{\rm s}$, one can solve for the ``cusp angle" $\theta_{\rm cusp}$ simply by solving $Q=U=0$ for $\theta_o$. In general, this entails solving three nonlinear equations: $Q(\alpha_{\rm cusp},\beta_{\rm cusp},\theta_{\rm cusp})=0$, $U(\alpha_{\rm cusp},\beta_{\rm cusp},\theta_{\rm cusp})=0$, and the geodesic equation. 
However, the calculation for purely vertical fields can be done analytically, and we present it here. Observe from \eqref{eq:kappadef} that $\kappa_1=\kappa_2=0$ for a purely vertical field only if $p^r\propto \mathcal{R}(r_{\rm s})=0$ and $f^{(r)}=0$. Expanding the fluid-frame components \eqref{eq:crossprod}, the latter condition is equivalent to $p^{(\phi)} = 0$. Solving $p^{(\phi)}(r_{\rm s},\lambda_{\rm cusp}, \eta_{\rm cusp})=0$ (with $\pm_s = -1$ for the direct image) and $\mathcal{R}(r_{\rm s},\lambda_{\rm cusp}, \eta_{\rm cusp})=0$ gives the conserved quantities $\lambda_{\rm cusp}$ and $\eta_{\rm cusp}$ at which the cusp will form. $\theta_{\rm cusp}$ is then found from the constraint that $\lambda_{\rm cusp}$ and $\eta_{\rm cusp}$ must be a solution to the geodesic equation for $r_{\rm s}$ and $\theta_{\rm s} = \frac{\pi}{2}$. Following \cite{Gralla2020a} (see (73) therein) the geodesic equation can be inverted to find:\begin{equation}
\label{eq:thetacusp}
\theta_{\rm cusp}=\arccos\left[\pm\sqrt{u_+}{\rm sn}\left(\sqrt{-u_-a^2}I_{r}^{\rm turn} \big|\frac{u_+}{u_-}\right)\right],
\end{equation}
where we have used $I_r^{\rm turn}$ \eqref{eq:Irturn} because by construction $p^r_{\rm s}(\lambda_{\rm cusp}, \eta_{\rm cusp})=0$, so the ray must begin at a radial turning point. To compare with (73) of \cite{Gralla2020a}, note that $\eta > 0$, $I_r^{\rm total}$ therein is $2 I_r^{\rm turn}$, and use \eqref{eq:intgeo} for $m=0$.\footnote{For the exact case of $a=0$, ~(\ref{eq:thetacusp}) reduces to\begin{align*}
    \theta_{\rm cusp}&=\arccos\left[\frac{\pm\sin\left(I_r^{\rm turn}\sqrt{\lambda_{\rm cusp}^2+\eta_{\rm cusp}}\right)}{\sqrt{1+\frac{\lambda_{\rm cusp}^2}{\eta_{\rm cusp}}}}\right].
\end{align*}}
We introduce the explicit $\pm$ in~\eqref{eq:thetacusp} to allow for $\theta_o>\pi/2$. Note that the $\pm$ indicates that for purely vertical fields, one loop in $Q,U$ space will exist for all observer inclinations that satisfy $\theta_{\rm cusp}<\theta_o<\pi-\theta_{\rm cusp}$. This follows additionally from Appendix~\ref{app:symmetries}, wherein we show that the polarization pattern for a purely vertical field will be reflected across the $\alpha$ axis upon taking $\theta_o\to\pi-\theta_o$.

More generally, fields with both vertical and equatorial components require a numerical solution for $\lambda_{\rm cusp},\eta_{\rm cusp}$, which can be substituted into (73) of \cite{Gralla2020a} to solve for $\theta_{\rm cusp}$ (now without the explicit $\pm$).

Figure~\ref{fig:cuspplot} illustrates some features of the direct image polarization corresponding to $\theta_{\rm cusp}$ for a sample magnetic field $\vec{B}=(0,1,1)$ around a Schwarzschild black hole, for which $\theta_{\rm cusp}$ must be computed numerically.  The left panel displays $\theta_{\rm cusp}$ as a function of emission radius $r_{\rm s}$, with dashed lines intersecting at $r_{\rm s} = 8$, $\theta_{\rm cusp} = 32.95^{\circ}$. The central panel displays the $Q,U$ loop at $r_{\rm s} = 8$, $\theta_{\rm cusp} = 32.95^{\circ}$, in which a cusp forms at the origin.  The right panel displays the EVPA as a function of orbital azimuth for observer inclinations of  $\theta_o=\theta_{\rm cusp} \pm 2^\circ$ in green and blue, respectively. As in Figure \ref{fig:qucompare2}, the EVPA rotates $2\pi$ for $\theta_o<\theta_{\rm cusp}$ and less than $2\pi$ for $\theta_o>\theta_{\rm cusp}$.

Finally, note that as the inclination continues to increase beyond $\theta_{\rm cusp}$, the lemniscate (figure-eight) that develops in the $Q,U$ curves and pulls through itself in Figure~\ref{fig:qupanelvert} at very large inclinations does not represent a singularity in EVPA. 

As remarked in Section~\ref{sec:Hotspots}, the location of the cusp can be shifted off the origin by including subimages (as suggested by Figure~\ref{fig:qusubimage}) and by Faraday conversion of Stokes $Q,U$ to circularly polarized Stokes $V$. It will also be shifted for hotspots with physical extent large enough for $\vec{p} \times \vec{B}$ to vary significantly over the emitting region. 

\bibliography{PolBib}{}
\bibliographystyle{utphys} 

\end{document}